**Multi-block chemometric approaches to the unsupervised spectral characterization of geological samples**

Beatriz Galindo-Prieto[1,2,*], Ian S. Mudway[1,2], Johan Linderholm[3,*], Paul Geladi[4]

[1] Environmental Research Group, School of Public Health, Faculty of Medicine, Imperial College London, London, UK

[2] MRC Centre for Environment and Health, School of Public Health, Faculty of Medicine, Imperial College London, London, UK

[3] Environmental Archaeology Laboratory, Dept. of Historical, Philosophical and Religious Studies, Umeå University, Umeå, Sweden

[4] Biomass Technology and Chemistry, Swedish University of Agricultural Sciences, Umeå, Sweden

[*] Corresponding authors (b.galindo-prieto@imperial.ac.uk, johan.linderholm@umu.se)

**ORCID's:**

Beatriz Galindo-Prieto: https://orcid.org/0000-0001-8776-8626

Ian S. Mudway: https://orcid.org/0000-0003-1239-5014

Johan Linderholm: https://orcid.org/0000-0001-7471-8195

Paul Geladi: https://orcid.org/0000-0001-6618-5931

## Abstract

As an example for the potential use of multi-block chemometric methods to provide improved unsupervised characterization of compositionally complex materials through the integration of multi-modal spectrometric data sets, we analysed spectral data derived from five field instruments (one XRF, two NIR, and two FT-Raman), collected on 76 bedrock samples of diverse composition. These data were analysed by single- and multi- block latent variable models, based on principal component analysis (PCA) and partial least squares (PLS). For the single-block approach, PCA and PLS models were generated; whilst hierarchical partial least squares (HPLS) regression was applied for the multi-block modelling. We also tested whether dimensionality reduction resulted in a more computationally efficient muti-block HPLS model with enhanced model interpretability and geological characterization power using the variable influence on projection (VIP) feature selection method.

The results showed differences in the characterization power of the five spectrometer data sets for the bedrock samples based on their mineral composition and geological properties; moreover, some spectroscopic techniques under-performed for distinguishing samples by composition. The multi-block HPLS and its VIP-strengthened model yielded a more complete unsupervised geological aggrupation of the samples in a single parsimonious model. We conclude that multi-block HPLS models are effective at combining multi-modal spectrometric data to provide a more comprehensive characterization of compositionally complex samples, and VIP can reduce HPLS model complexity, while increasing its data interpretability. These approaches have been applied here to a geological data set, but are amenable to a broad range of applications across chemical and biomedical disciplines.

## 1. Introduction

Grouping samples by their properties using multivariate latent models based on spectral data is usual in fields related to natural sciences and medicine [1–3]. In recent years, chemometric methods have become popular to integrate and analyse multi-modal data and numerous algorithms have been developed [4,5]. However, the fusion of different types of spectral data sets (e.g., NIR, Raman, and XRF) in a unique and parsimonious multi-block model with the purpose of enhancing pattern recognition, characterization, and grouping of compositionally heterogeneous geological specimens has not been achieved. Some barriers to the adoption of these techniques lie in the high computation power required, the necessary statistical and machine learning expertise, and the need for easy-to-interpret visualizations of the underlying patterns and structures connecting observations across data sets. In disciplines such as geology or archaeology, due to the explorative and discovery nature of the sample collection during the field work, there is an interest in using multivariate models (e.g., PLS) that do not require the inclusion of sample classes or compositional categories prior to statistical modelling, rather than their discriminant analysis versions (e.g., PLS-DA) that use class/categorical knowledge provided in advance.. In this paper, we test a multi-block modelling approach based on partial least squares, combined with model dimensionality reduction using variable influence on projection (VIP), as a potential methodology for an improved unsupervised characterization based on the analysis of 76 bedrock specimens. We would like to emphasize that we aim to explore the possibilities of PLS for achieving a natural grouping meeting in the middle of pattern recognition and formal classification in the fields of archaelogy and geology with the ultimate purpose of testing whether PLS multi-block approaches can perform better than their single-block analogous approaches for geological sample grouping/characterization. It is out of the scope of this paper to perform a formal classification (e.g., using soft independent modelling of class analogy) or prediction. Therefore, to avoid confusion, we will not refer to any observed grouping of samples in the results of the models as “classification”, since the PLS models in this work do not return sample class labels, being the approach adopted in this paper a natural-/human- interpretation of the score plots generated in the chemometric latent models. In addition, by definition, any regression model is "supervised" due to the rotation in the hyperspace of the latent variables to maximize the covariance between the **X** and **Y** data matrices since this rotation uses as reference **Y** (thereby, **Y** "supervises" the rotation); however, in this paper, we will use the terms "supervised" and "unsupervised" in the sense of existence or absence of pre-defined classes [6] rather than an indication of existence or absence of adjustment of **X** to **Y**. So, the term "unsupervised" is here used in the sense of using models that have not being set up with information about the classes/categories of samples in advance, i.e., to differentiate our PLS modeling approach without pre-defined classes from any other PLS discriminant approach with pre-defined classes (initial categorical information) in the model set-up (as it happens in PLS-DA).

There are three main types of bedrock: igneous/magmatic, sedimentary, and metamorphic. Frequently, bedrock matrices are heterogenous, and sampling them may be challenging for most analytical techniques. In this study, data sets from five field instruments were used to investigate the use of multi-spectral approaches to the characterization of complex samples; more specifically, two near-infrared, two Fourier transform Raman, and one X-ray fluorescence data sets. X-ray fluorescence (XRF) [7] can be used to detect and quantify chemical elements from Mg (atomic number Z=12) to U (Z=92). XRF instrumentation can be made portable, battery driven, and therefore is often used in geological and archaeological field applications [8–10]. XRF spectra and elemental concentrations were measured in 76 solid rock samples; and, additionally, two portable near infrared (NIR) and two portable FT-Raman instruments were used with the same samples. In the geological and archaeological fields, there is increasing interest in whether NIR [11] and FT-Raman [12] spectra can also be used to classify rock specimens, rather than just using XRF (which cannot detect, for example, carbon related samples), as well as whether a combination of the different types of spectra (NIR, FT-Raman, and XRF) could lead to a better discrimination of geological specimens. Some authors have described the use of FT-Raman [13,14] and NIR spectra [15,16] for geological samples.

All spectral data sets were derived from the 76 geological samples with a varying number of variables (measurements). The data consisted of 42 variables for XRF (element concentrations), 603 and 1451 for two FT-Raman instruments (wavenumbers), and 128 and 1501 for two NIR instruments (wavelengths). After pre-processing of the raw data, the data sets provide an opportunity of testing different types of multivariate models and their interpretations. We show and compare, in terms of data interpretation and geological characterization, the results of analysing the five spectral data sets; firstly, separately by inspection of single-block PCA [17] and PLS [18,19] models, and afterwards, combined by inspecting multi-block HPLS models [20]. A post-modelling VIP variable selection [21,22] was carried out to reduce the dimensions of the hierarchical partial least squares (HPLS) regression model and improve its interpretability of the associations between spectroscopy type and geological sample group.

## 2. Materials and Methods

In this section, a description of the data sets and the instruments used to generate them is provided, as well as a brief explanation of the methodologies and algorithms employed for generating the multivariate and multi-block models.

### 2.1. Data sets and instrumentation

The multi-block data set consists of five spectral data matrices derived from the same 76 geological bedrock samples measured using five different instruments (**Fig. 1**). These inorganic samples have relevance for geological and archaeological studies. All originated from Europe, mainly from Sweden. All samples were roughly palm size and classified by type by geologists. This collection of samples is not meant to be a global selection but was collected as a didactical tool for archaeology-geology students. The selection of rocks covers all main types: igneous/magmatic, metamorphic, and sedimentary, being heterogeneous with regard to structure and mineral composition, and have also uneven surfaces. Of particular interest was the inclusion of various quartzes and quartzites as they were important materials in prehistoric tool production [23]. Categorical information for each sample is provided in **Tables S1-S2** of **Supporting Information 1**, which includes the general and the given sample names, general classification, the base mineral, additional mineral information, chemical composition, and the sampling location (site, province, and country).

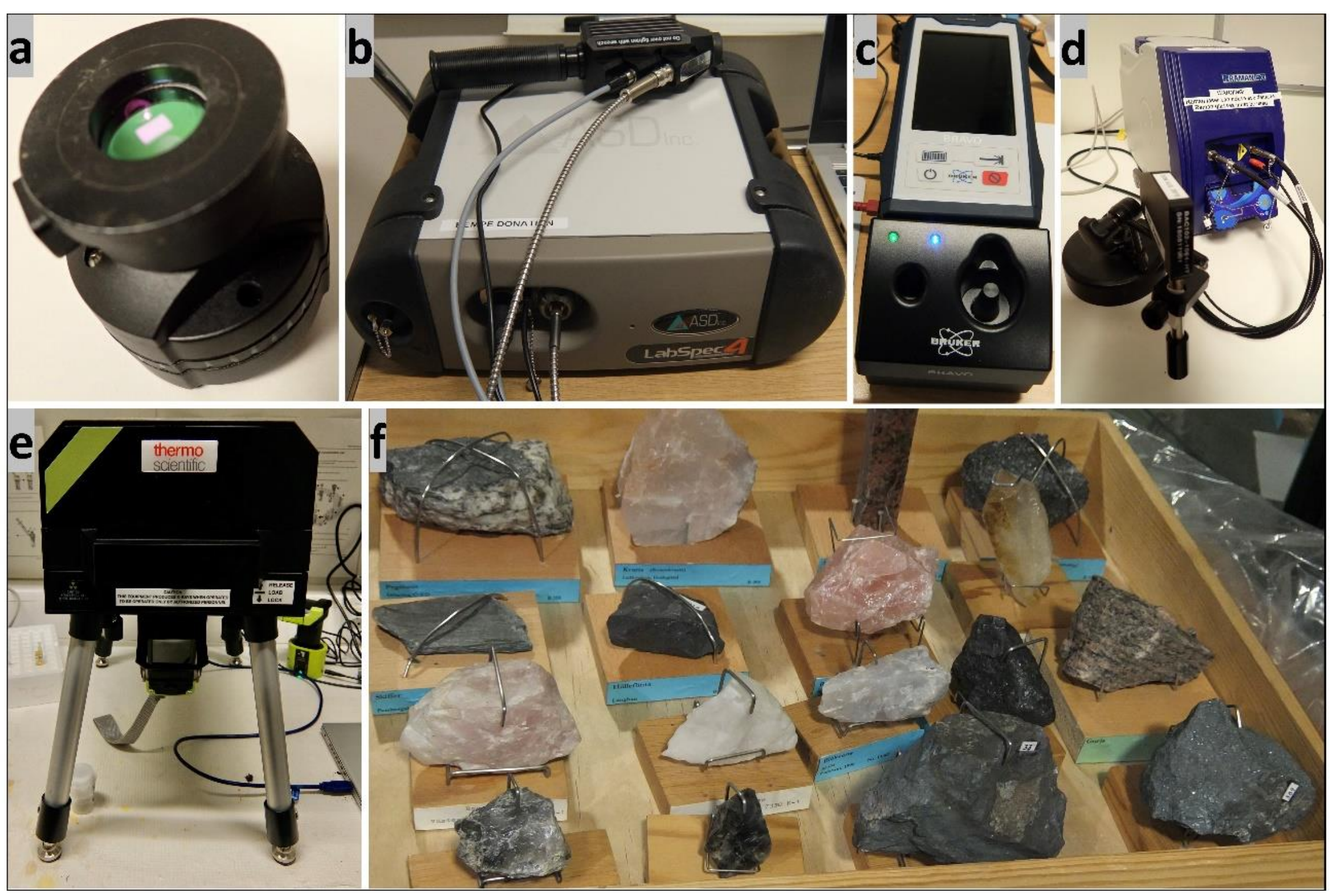

***Fig. 1*** *Pictures of the five spectrometers, i.e., (a) Micro-NIR, (b) ASD-NIR, (c) Bruker-Raman, (d) i-Raman, and (e) XRF; and (f) some representative specimens.*

A description of the contents and dimensions of the five data sets is given in

**Table *1***. All used instrumentation was field-adapted and had contact probes and internal illumination. The first NIR spectrometer was an Analytical Spectral Device (ASD) LabSpec 4

(with range of 350:1:2500 nm, large wavelength range including UV and VIS, and high spectral resolution) with one Si and

***Table I*** *Description of the five raw data sets, including their original reference in the literature and applied preprocessing.*

| Instrument and manufacturer | Variables obtained | Data matrix | Variables used | Lit Ref. | Preprocessing |
|---|---|---|---|---|---|
| **ASD LabSpec4 - Malvern Panalytical** | Wavelengths 350:1:2500 nm | 76x1501 | 1000-2500 nm | [31] | SNV + MC |
| **Bruker BRAVO FT-Raman** | Dual laser 785 and 852 nm, wavenumbers 300:2:3200 $cm^{-1}$ | 76x1451 | All | | MSC + MC |
| **FT iRaman - Metrohm** | Laser 1064 nm, wavenumbers 92:4:2507 $cm^{-1}$ | 76x603 | All | | MSC + MC |
| **VIAVI MicroNIR** | Wavelengths 908:6:1676 nm | 76x128 | 1000-1676 nm | [32,33] | SNV + MC |
| **EDXRF Thermo Scientific Niton XL5 Plus** | Quantified elemental concentrations 42 | 76x42 | 16 (Bal, Mg, Al, Si, P, S, Cl, K, Ca, Ti, Mn, Fe, Zn, As, Ba, Pb) | | Pareto + MC |

*SNV stands for standard normal variate, MC for mean-centrering, and MSC for multiplicative scatter correction. A column showing the finally used for data analysis wavelengths, wavenumbers, and elements (ordered by atomic number), is also included.*

two cooled InGaAs arrays as detectors; whilst the second NIR spectrometer was a compact VIAVI Micro-NIR (with range 908:6:1676 nm, wavelength range covering a specific part of the NIR, and lower spectral resolution than the ASD LabSepc 4) with a InGaAs array as detector. The third and fourth data sets were obtained from two Raman spectrometers: a Bruker BRAVO portable FT-Raman spectrometer (with laser ca 700 nm, and range 300:2:3200 $cm^{-1}$) and a FT i-Raman field instrument (with a 1064 nm excitation laser, a range of 98:4:2507 $cm^{-1}$, and fibreoptic probe). The fifth data set was produced by a XRF instrument (EDXRF, Energy Dispersive Thermo Scientific Niton XL5 Plus) again configured for field use, which provides elemental concentration data (from Mg to U), using a 5W Ag anode X-Ray tube and Silicon drift detector (spot size: 8 nm); the instrument calibration for mining mode was used for quantification. Whilst field instruments were used for all measurements, for this study the geological samples were measured under controlled laboratory conditions in a dark room to minimize stray light. In addition, as the samples were not powders, but solid bedrock specimens, and all five instruments used probes of differing diameters (**Fig. S1** of **Supporting Information 1**), five replicate measurements were taken on each sample and the average of the replicates used for each specimen.

### 2.2. Data pre-processing strategy

**Fig. S2-S6** of **Supporting Information 1** show the raw and the pre-processed spectral data sets. All pre-processing was done with in-house MATLAB code (version R2023a, The MathWorks, Natick, MA, USA), except the multiplicative scatter correction that was done using R (version 4.3.1, R Core Team, Vienna, Austria). All visualizations shown in **Fig. S2-S6** were obtained using in-house MATLAB code.

The raw XRF data set initially contained 42 variables. All variable names correspond to elements of the periodic table but *Bal* (the *Balance* variable) that represents all elements below Mg (i.e., with $Z < 12$) that were considered *non-determined matter* by the XRF instrument. Three XRF variables (Hf, Re and Ta) had their values *non-detected* (ND values) for all samples, so these variables (elements) were discarded. For the rest of the XRF variables, where the element was not detectable null concentrations were imputed rather than the limit of detection (LOD). The imputed data set was mean-centred and Pareto-scaled.

The ASD-NIR and Micro-NIR data matrices were standard normal variate (SNV) transformed [24] to remove the multiplicative interferences of scatter and particle size, and afterwards, mean-centred. Sample 68 was removed from the Micro-NIR data set since the instrument could not produce any measurement due to the very dark colour of the specimen. This paper aims to evaluate NIR and FT-Raman spectroscopy, with XRF as response, for use in multivariate models for distinguishing geological samples without any other supportive information or data; therefore, UV and VIS wavelengths (*ca.* < 1000 nm) of the acquired NIR spectra were removed before starting the data analysis.

Both FT-Raman data sets were treated for non-linear scatter-effects by means of multiplicative scatter correction (MSC) [25], and afterwards, mean-centred. Some samples of the Bruker- and i- Raman data sets (sample 03, and samples 35 and 37, respectively) were identified as outliers by direct spectra inspection. Afterwards, the FT-Raman data were inspected by PCA score plots resulting in the identification of samples 53 and 73 for Bruker-Raman, and samples 26, 39, 42 and 62 for i-Raman, as outliers.

### 2.3. Elemental and mineral composition description

The characteristics and composition of the rock specimens are summarised in **Tables S1-S2**. Their basic geology (magmatic, metamorphic, or sedimentary) and their basic mineral composition give them unique properties that can be differentiated in the latent structures of multivariate latent variable models (e.g., in the principal components of PCA models). To visualize these differences in the elemental composition and properties of the samples, a PCA model of the XRF data was generated as outlined below. The selected visualization to show the different composition and properties of the samples was a biplot, where both scores and loadings of the PCA model can be inspected and interpreted. The combination of scores and loadings in the same visual representation makes it possible to discriminate the clusters of samples that are similar, as well as the elements related to each cluster (i.e., to each group of samples with similar composition and geological properties), see **Fig. 2**.

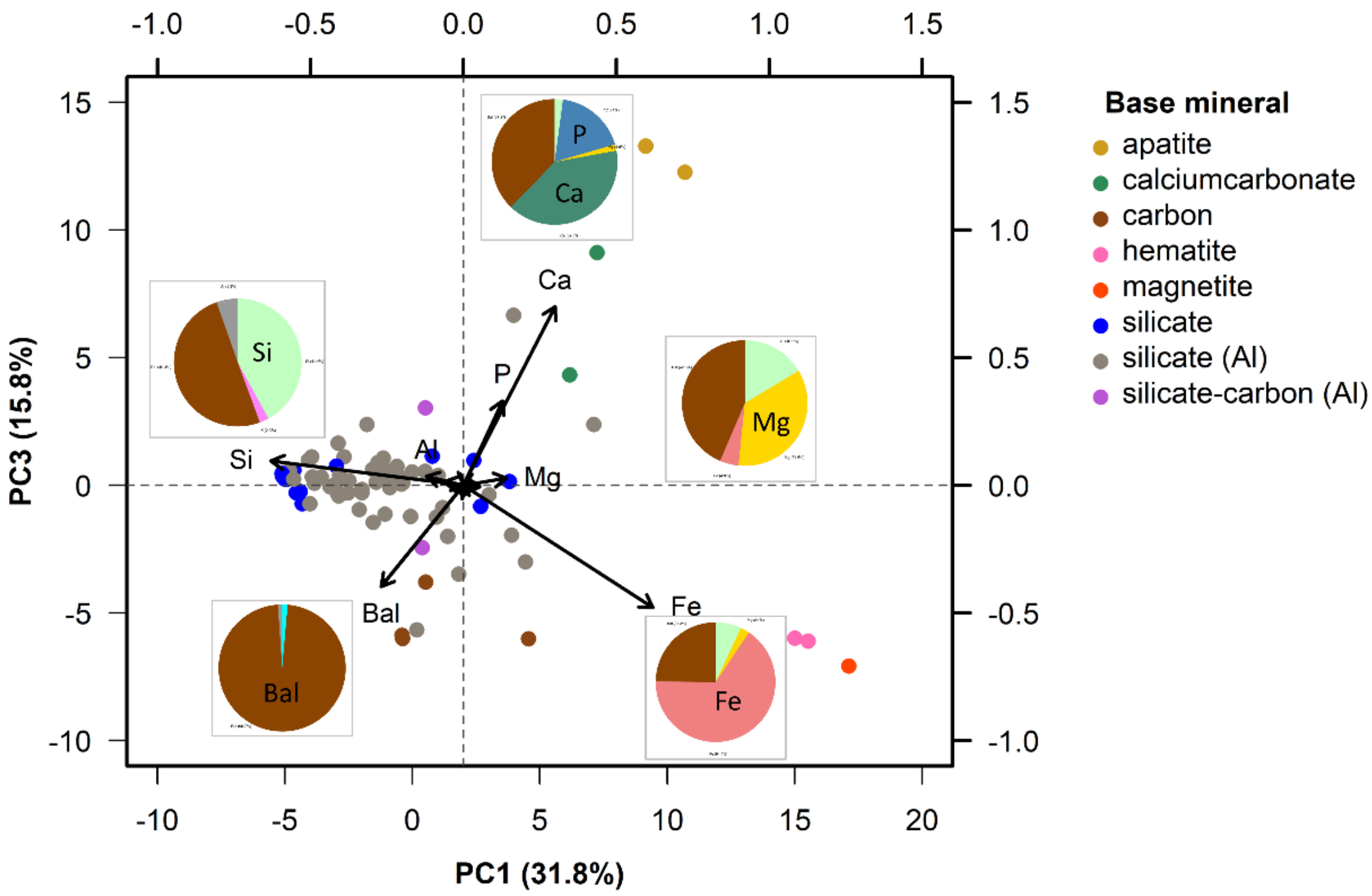

***Fig. 2*** *Biplot of a PCA model of the XRF data. The legend shows the colour used for each type of sample in relation to its base mineral. The X bottom axis and the Y left axis provide the score values for the first and third principal components (PC1 and PC3). The X superior axis and the Y right axis provide the loading values for PC1 and PC3 respectively. Points represent the samples, and arrows represent the variables (elements); only the most informative variables (Bal, Si, P, Ca, Fe, Mg and Al) have been labelled, the rest (arrows for S, Cl, K, Ti, Mn, Zn, As, Ba and Pb, located at the coordinates centre of the loadings) have not been labelled for better readability of the figure.*

### 2.4. Multivariate and multi-block statistical methods

Single-block and multi-block models based on partial least squares are constructed in this paper. The results of the single-block and multi-block multivariate latent models will be shown as statistics of the models and visualizations of the scores ($t_a$) and loadings ($p_a$) obtained for each latent structure (model component, $a$). The PCA model of the XRF data was computed and visualized using R (version 4.3.1). The PLS and HPLS models were calculated using MATLAB (version R2023a), and their visualizations were obtained utilizing either R (version 4.3.1) or MATLAB (version R2023a). VIP was calculated and visualized using MATLAB (version R2023a).

## 3. Results

### 3.1. Sample characterization by XRF and Principal Component Analysis

**Fig. 2** illustrates the differences in elemental composition and properties of the 76 bedrock samples in a biplot of a PCA model originated from the XRF data set. The first principal component explained a 31.8% of variation, the second a 25.1%, the third a 15.8%, the fourth a 11.2% and the fifth a 9.0%. The biplot of **Fig. 2** shows the relationships between samples (grouping of bedrock specimens), between variables (XRF elements, represented by black arrows), and between samples (specimens) and variables (elements). The sample points were coloured by base mineral, i.e., apatite, calcium-carbonate, carbon, hematite, magnetite, silicate, silicate with aluminium (Al), and silicate-carbon (which also contains Al). The black loading arrows point to the direction in which certain groups of samples separate from the others according to their elemental composition. Pie charts of the elemental composition for each one of the 76 bedrock specimens are available in **Supporting Information 2**; a few of them were inserted next to certain groups of samples in **Fig. 2** to provide a clearer view of their elemental composition. For instance, the biplot showed that hematite and magnetite specimens, compositionally rich in iron (Fe), clustered on the bottom-right corner of the biplot where the Fe variable was also located; whilst the apatite samples, enriched with calcium (Ca) and phosphorus (P), were clustered on the top of the biplot. It is worth noting that the inclusion of pie charts in the study provides additional external validation to the PCA model pattern recognition generated from the XRF data. As it can be seen in **Fig. 2**, the PCA model identified Ca, Si, Bal (all elements with $Z < 12$) and Fe as the more informative variables for unsupervised sample characterization. The first principal component (PC1) separated the specimens that contain mainly Bal and Si from the specimens that contain mainly Fe and Mg. Apatite samples have positive high score values for PC3; whilst hematite, magnetite and carbon samples have negative score values. The calcium-carbonate samples are located in the middle of the apatite and carbon clusters. We would like to clarify that the biplot of PC1 and PC2 was inspected and also showed group separation; however, the use of PC1 and PC3 for the biplot seemed to show a few groups slightly clearer. In addition, the use of PC1 and PC3 for **Fig. 2** was also preferred because the spatial distribution of the scores in the plot made easier the insertion of the pie charts in the figure.

### 3.2. Single-block PLS models

Individual PLS models of the NIR and FT-Raman data sets, using the XRF data as response matrix, were generated. All models were leave-one-out cross-validated, and the number of optimal model components for each model was determined based on the values of root mean square error (RMSE) per latent variable and the total amount of variation explained by the model. Model cross-validation ensured a proper extraction of latent structures for obtainment of informative scores for furhter analysis; i.e., model validation helped to achieve model stability towards known and unknown sources of variation [26]. The PLS models were built using a reduced **Y-**block consisting of 16 (out of 42) XRF elemental variables; the 16 elements were selected based on their relative abundance in the Earth's crust and their relevance for environmental science. Sample 68 was excluded from the Micro-NIR PLS model because of having all its measurements missing as its dark colour made measurement with the Micro-NIR spectrometer impossible. In the FT-Raman PLS models, based on outliers' inspection, samples 03, 53 and 73 were excluded from the Bruker-Raman model; and samples 26, 35, 37, 39, 42 and 62 from the i-Raman model.

A 11-component ASD-NIR PLS model and a 7-component Micro-NIR PLS model were computed. The ASD-NIR and the Micro-NIR PLS models explained a 99.4% and a 99.7% of the total **X-**variation respectively. **Tables S3-S4** (in **Supporting Information 1**) provide the values of RMSE and **X-** and **Y-** explained variation of the ASD-NIR and Micro-NIR PLS models for each model component, as well as the cumulated total explained variation from first to last latent variable extraction. **Fig. 3** represents the scores of the first two latent variables for the ASD-NIR and the Micro-NIR PLS models, the sample points were coloured by base mineral and given a specific shape according to their basic geology. For ASD-NIR, **Fig. 3a** showed a clear cluster of magmatic silicates (blue circles) with low LV1 score values, whilst metamorphic carbon specimens (brown crosses) were clustered with high LV1 score values. Hematites (pink circles) were clearly grouped showing different mineral and geological properties. Silicate-carbon and calcium-carbonate samples were also clearly clustered, but not separated since their mineral and geological properties are more similar to the rest of samples than in the hematite samples case. Almost all specimens with both metamorphic and sedimentary properties (squared crosses) were clustered. However, apatites did not group as expected from their geological and mineral properties, and the magnetite sample was clearly separated from the hematite samples despite similar elemental composition.

Micro-NIR (**Fig. 3b**) also showed clustering for almost all magmatic silicate samples, and for all calcium-carbonate and carbon samples. However, this was not the case for the apatite and hematite specimens; furthermore, the first five model components (LV1-LV5) were inspected in the Micro-NIR PLS model, and none of them separated apatite and hematite samples. We would like to emphasize that although the score plots were visualized for all components of all the models of this paper, only the figures related to LV1 and LV2 are shown for the sake of succinctness. Besides, the first two components were the most relevant ones for sample clustering.

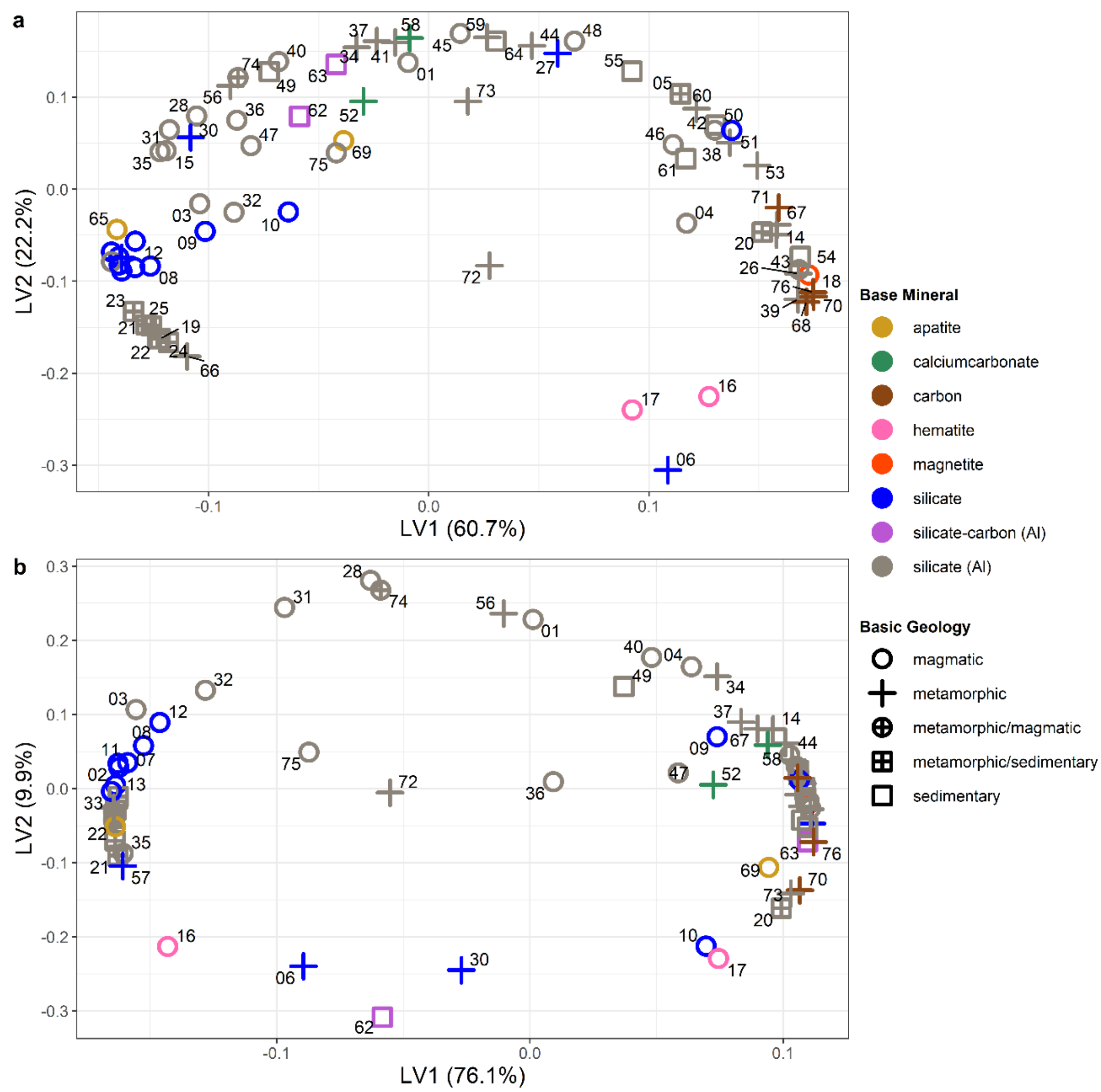

***Fig. 3*** *PLS scores for LV1 and LV2 for the (**a**) ASD-NIR and (**b**) Micro-NIR models. Samples are represented by points coloured according to base mineral and shaped according to basic geology. Magmatic specimens are represented by circles, metamorphic by crosses, metamorphic/magmatic by circled crosses, sedimentary by squares, and metamorphic/sedimentary by squared crosses.*

The models' outputs not shown here (such as scatter plots of scores, loadings, and biplots; as well as some relevant statistics) are available in the repository of **Supporting Information 2**.

Similar to NIR, a 15-component Bruker-Raman PLS model and a 7-component i-Raman PLS model were computed. The Bruker- and the i- Raman PLS models explained a 87.8% and a 97.4% of the total **X-**variation respectively. The score plots for the first two model components (LV1 and LV2) are shown in **Fig. 4** (and the loading plots for LV1 and LV2 are shown in **Fig. S7** of **Supporting Information 1**). Tables of RMSE and explained variation for **X** and **Y** of each PLS model are provided in **Supporting Information 1** (**Tables S5-S6**), the tables provide the statistics for each model component and the cumulated total explained variation from first to last latent variable extraction. The scores represented in **Fig. 4a** showed that Bruker-Raman clusters magmatic silicates better than i-Raman (**Fig. 4b**); however, i-Raman seems to capture the similarities between the two apatites better (resulting in a more defined cluster). Calcium-carbonate and sedimentary silicate-carbon samples were well classified by both FT-Raman techniques. Interestingly, unlike ASD- and Micro- NIR, only Bruker-Raman was able to differentiate between metamorphic carbon specimens with very high content (*ca.* 96%) of *Bal* (samples 68 and 71) and metamorphic carbon specimens with not so abundant amounts of *Bal* (*ca.* 72%) and significant but minor presence of other elements; **Fig. S8** (**Supporting Information 1)** shows the elemental composition of the specimens of the metamorphic carbon group.

A variable importance on projection (VIP) [21] assessment was performed for each PLS model (all VIP plots are available in **Supporting Information 2**) to determine which were the most important wavelength and wavenumber variables for sample characterization. As example, we show the VIP plot for the Bruker-Raman PLS model (that uses the XRF data as response) in **Fig. S9**, **Supporting Information 1**. The wavenumber variables with VIP > 1 *a.u.* (i.e., above the threshold red line of **Fig. S9**) are the most contributing variables for unsupervised characterization of the specimens. The biplot shown in **Fig. S10** is a visualization of the associations between the geological samples and some of the most relevant variables (wavenumbers) for the characterization of the Bruker-Raman data using partial least squares regression. The biplot clearly shows that LV4 explained the slates (giving them high scoring, which located them on the top of the plot) and the variables that helped to explain the slates were ca. 402-412 $cm^{-1}$. The calcium-carbonates were explained in the range of 1086-1098 $cm^{-1}$, and LV4 separated them from the rest by giving them very negative score values. LV2 separated (giving low scores) the apatite (together with the quartz feldspar), see left side of the biplot (**Fig. S10**), which was distinguished by the first wavenumbers of the Bruker-Raman spectra. **Fig. S9-S10** show evidence that VIP highlights the most important regions of the spectrum for the characterization and grouping of geological samples.

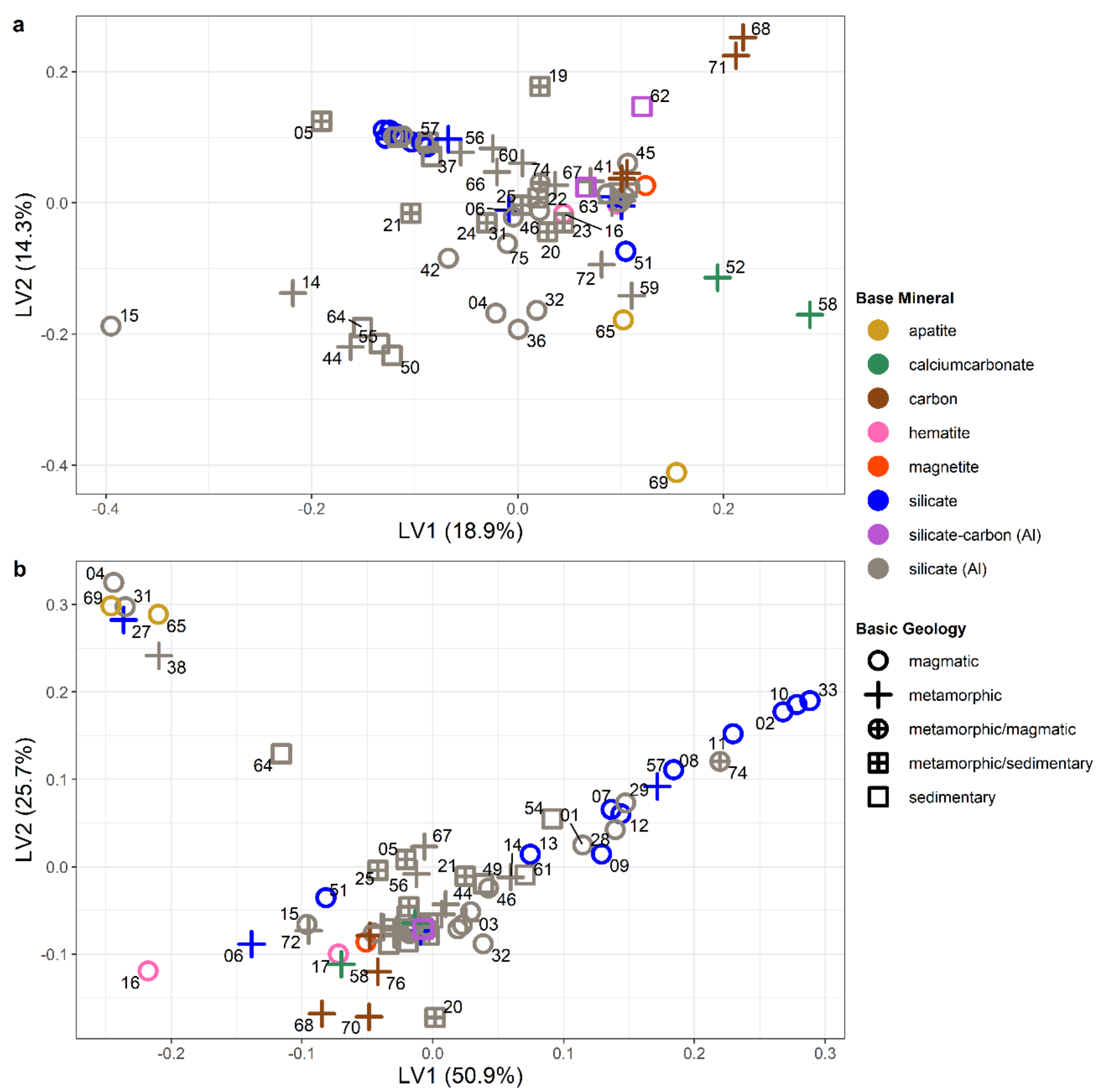

***Fig. 4*** *PLS scores for LV1 and LV2 for the (**a**) Bruker-Raman and (**b**) i-Raman models. Samples are represented by points with colour according to base mineral and shape according to basic geology. Magmatic specimens are represented by circles, metamorphic by crosses, metamorphic/magmatic by circled crosses, sedimentary by squares, and metamorphic/sedimentary by squared crosses.*

### 3.3. Multi-block HPLS model using NIR, FT-Raman and XRF data

To evaluate whether the two NIR and two FT-Raman data sets could support each other to yield a latent model with better interpretability and higher characterization/grouping power, a multi-spectra HPLS model (i.e., including NIR, FT-Raman and XRF) was built using as super **X-**matrix the 38 **X-**scores of four sub-level individual (single-block) PLS models, with only 66 of the 76 samples (i.e., excluding the outliers 03, 26, 35, 37, 39, 42, 53, 62, 68 and 73), and as super **Y-**matrix the corresponding **Y-**scores (with the purpose of summarizing the overall structure of the XRF data). Since the scores obtained from the PLS models came from different spectrometers, they were scaled prior to multi-block modelling. The HPLS model was leave-one-out cross-validated, and 14 latent structures (model components) were extracted. The number of optimal model components was determined according to RMSE and explained variation values for each latent variable. The HPLS model explained a 73.1% of the total **X-**variation and an 85.4% of the total **Y-**variation. **Table S7** shows the RMSE values and the **X-** and **Y-** explained variation, which comes from both the NIR and FT-Raman spectra (so, the multi-block model fuses the information contained in the latent structures of the four spectra, i.e., the two NIR and the two FT-Raman), for each HPLS model component. As a sensitivity analysis, the HPLS modelling was repeated using the scores of sub-level individual (single-block) PCA models to determine whether a better grouping could be achieved; the results (**Fig. S11** in **Supporting Information 1**) were not significantly better than using the scores of sub-level individual PLS models.

The 14 super-scores and super-loadings obtained after running the HPLS model were inspected. **Fig. 5** shows the HPLS super-scores plot for the first two model components (LV1 and LV2), that come from the **X-** and **Y-** scores of the four individual, single-block PLS models (ASD-NIR vs XRF, Micro-NIR vs XRF, Bruker-Raman vs XRF, i-Raman vs XRF). The HPLS super-loadings scatter plot for LV1 and LV2 is shown in **Fig. S12a**. The multi-block approach yielded a cleaner grouping of the samples according to base mineral and basic geology than the score plots previously inspected for each original single-block PLS model (**Fig. 3-4**). The HPLS model, where both NIR and FT-Raman variances complemented each other, provided a complete characterization of the geological samples in one unique model. Magmatic silicates, hematites, calcium-carbonates, and metamorphic carbon specimens were clearly grouped in **Fig. 5**. Due to the removal of sample 68, the ability of the model to separate the two types of carbon samples could not be assessed. Sedimentary and magmatic bedrocks formed more concentrated clusters than metamorphic, and LV1 separated non-silicate samples (apatites, calcium-carbonates, hematites, magnetites) from the three groups of silicates. LV2 separated hematites (that contain more than 64% of Fe, see **Fig. S13**) from apatites (**Fig. S14**) and calcium-carbonates (**Fig. S15**), with high content of Ca, and without Fe (except the apatite sample 65 that, due to its 7.7% of Fe, was assigned lower LV2 scores than the other samples in the same group). The carbon samples (**Fig. S8**), interestingly, also separated according to their elemental composition; having sample 70 (the only one that contains Fe) lower LV2 score values than samples 70 and 76. Therefore, even if there may be more chemical interactions that could be discover by looking at this and other score plots provided by the HPLS model, it seems reasonable to think that LV1 explains the content of Si in the samples, and LV2 explains the content of Fe in them. This interpretation is supported by the fact that Si and Fe are two of the strongest responses in the multi-block model.

It is worth noting that the score values of LV2 also increase when Ca content increases in the samples.

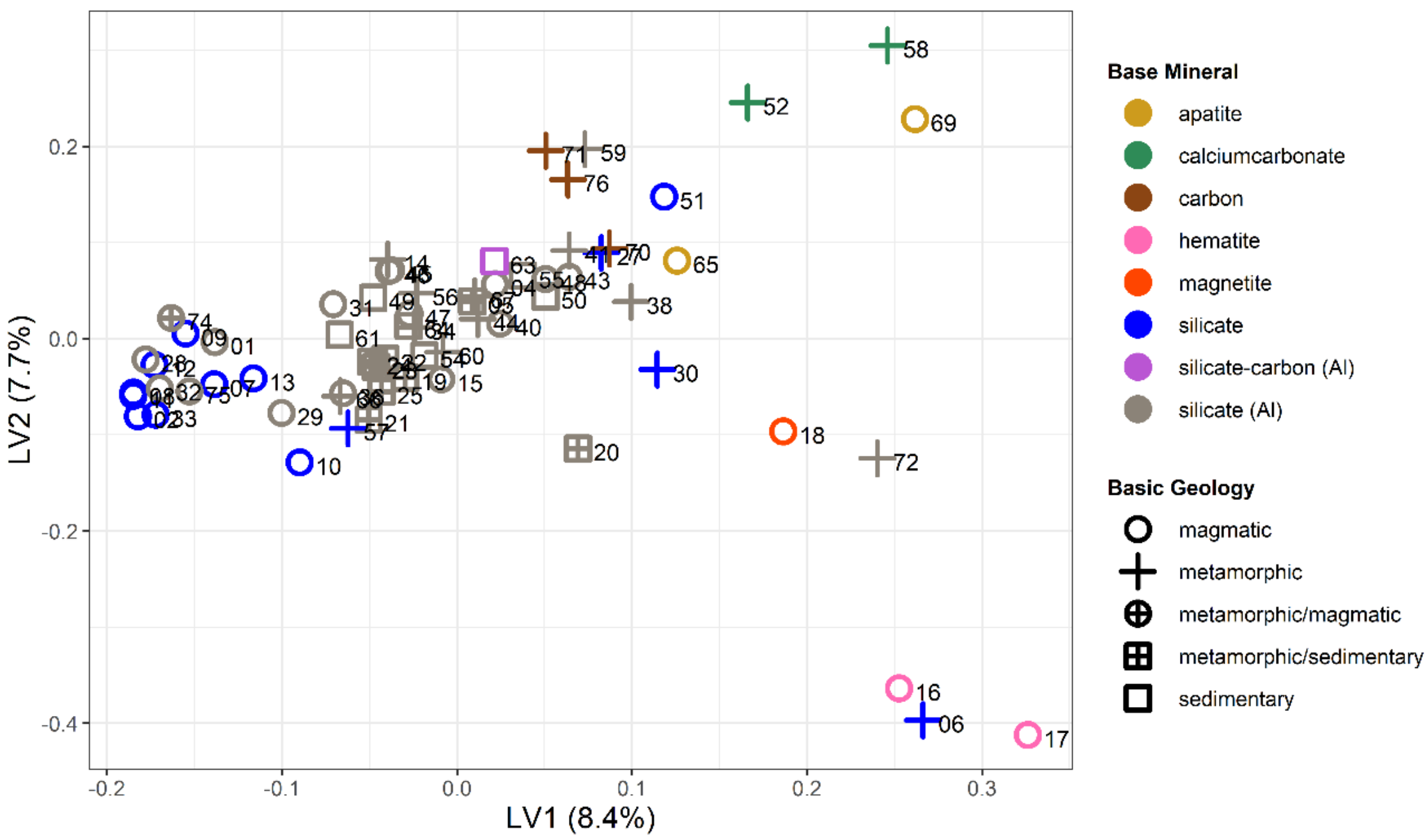

***Fig. 5*** *Super-scores plot of a 14-component HPLS model using 38 PLS scores from the four individual PLS models. The first model component (LV1) is represented in the X-axis, and the second (LV2) in the Y-axis. The legends indicate the colour for each base mineral category, and the shape for each basic geology type.*

## 3.4. Dimensionally reduced VIP-HPLS model

The variable importance on projection (VIP) method was used to select the most relevant PLS score-variables from the first HPLS model. VIP selected 14 out of the 38 PLS scores (**X-**variables of the HPLS model) as important for the multi-block model (the 14 score-variables are named in the **Y-**axis of **Fig. 6**). Variables that had a VIP value higher than 1 *a.u.* (red vertical line in **Fig. 6**) were identified as important for model interpretation and sample characterization / grouping.

A new HPLS model was generated using only the 14 PLS score-variables selected by VIP from the first HPLS model. This second model was significantly stronger than the first HPLS model explaining all (100%) the **X-**variation of the NIR and FT-Raman data, however it explained slightly less **Y-**variation (68.7%) of the XRF data. The explained variation (total and per model component) of the HPLS model dimensionally reduced by VIP is shown in **Table S8** of **Supporting Information 1**. From the inspection of the first two super-scores of the VIP-reduced HPLS model, it was noticed that they explained more variation (19.6% and 17.0%) than their analogous model components in the first HPLS model (which explained 8.4% and 7.7%). **Fig. 7** shows the super-scores for LV1 and LV2 of the VIP-reduced HPLS model (the corresponding loadings are shown in **Fig. S12b**, **Supporting Information 1**). In terms of sample grouping according to base mineral and basic geology, the results were similar to the first HPLS; however, the VIP-reduced model separated the magmatic silicates with Al (represented by grey cercles in **Fig. 7**) from the ones without Al (blue circles) better than the first HPLS model (**Fig. 5**). Besides, less variables (14 PLS scores instead of 38) explained more NIR and FT-Raman spectral variation, in a stronger and a more parsimonious VIP-refined HPLS model.

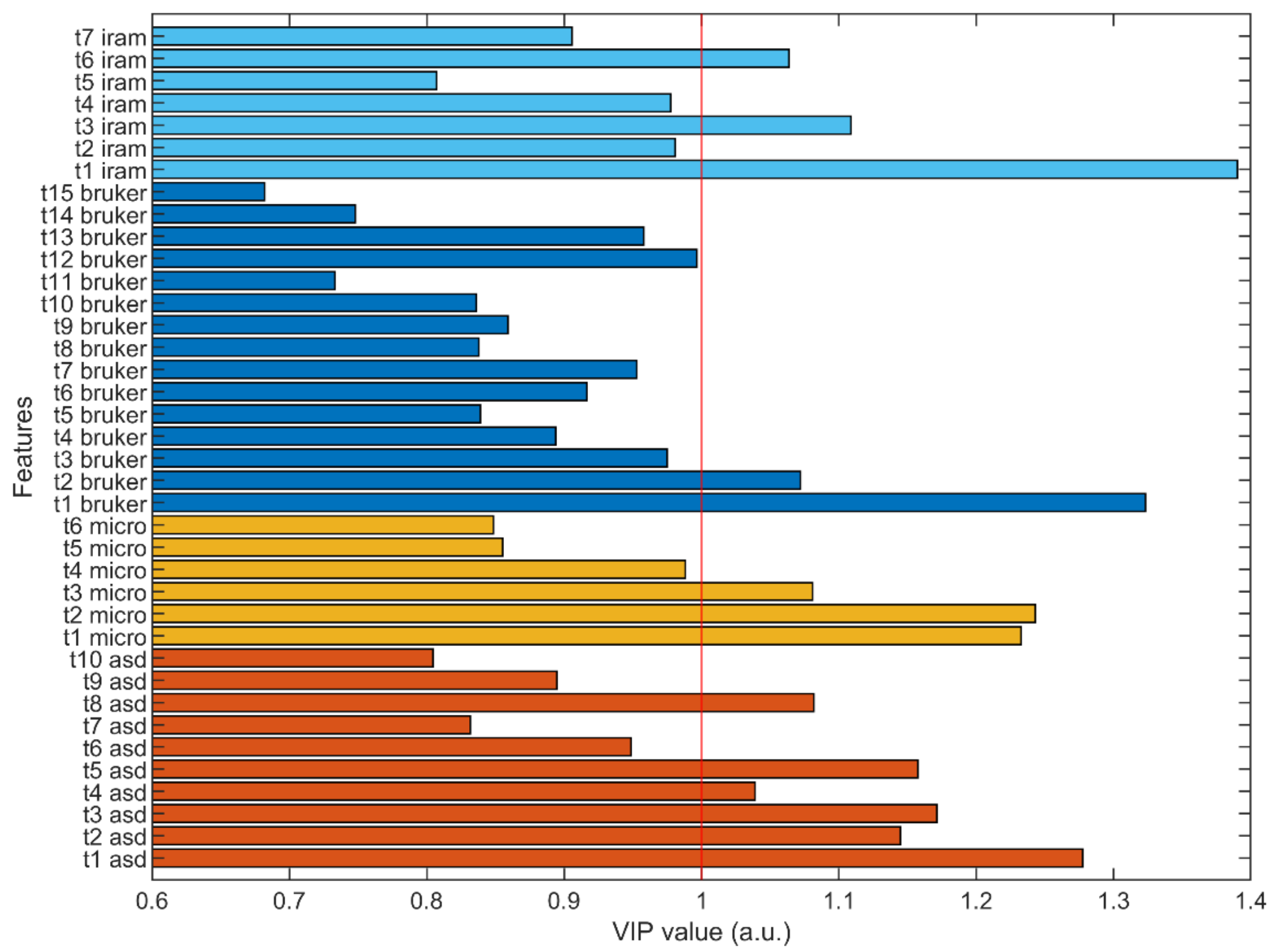

***Fig. 6*** *Variable influence on projection (VIP) bar plot for the HPLS model generated from the 38 PLS score-variables of the four individual PLS models. The 38 score-variables are indicated as features in the Y-axis, and the corresponding VIP values are indicated in the X-axis. A red vertical line marks the threshold for importance at VIP=1. The colours of the bars indicate which data set each feature (score-variable) comes from (ASD-NIR in red, micro-NIR in yellow, Bruker-Raman in dark blue, and i-Raman in light blue).*

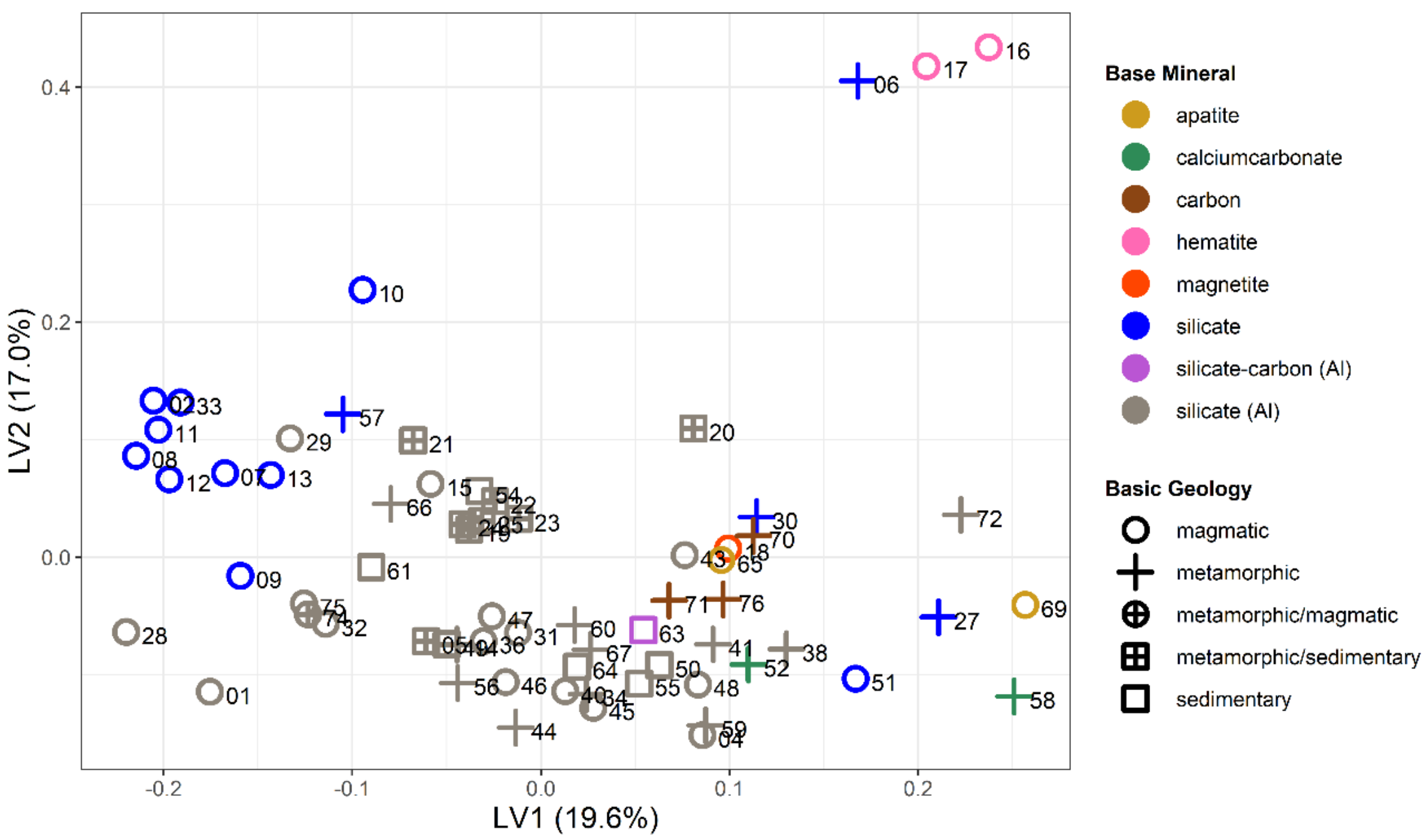

*Fig. 7 Super-scores plot of a 14-component HPLS model using only 14 PLS scores from the four individual PLS models (ASD-NIR vs XRF, Micro-NIR vs XRF, Bruker-Raman vs XRF, i-Raman vs XRF). The first model component (LV1) is represented in the X-axis, and the second (LV2) in the Y-axis. The legends indicate the colour of each base mineral category, and the shape of each basic geology type.*

In terms of model interpretation and model ability to explain the variance coming from the different spectroscopic techniques, a summary of the number of model components (*a.k.a.* latent variables, LV), percentage of explained **X-**variation ($R^2X$), and percentage of explained **Y-**variation ($R^2Y$), for each model is shown in **Table II**. Both NIR PLS models explained more than 99% of the total variation contained in the NIR spectra. For FT-Raman, the i-Raman PLS model explained more total variation (97.4%) contained in the spectra than the Bruker-PLS model (87.8%). For the multi-block HPLS models, the reduction of dimensionality by applying VIP variable selection highly impacted the interpretability of the model. The HPLS and the VIP-refined HPLS models were both built with the same number of components; hence, they are fully comparable. The first HPLS explained a 73.1% of NIR and FT-Raman spectral variation using 38 variables (PLS scores); however, the VIP-refined HPLS model explained all (100%) of the NIR and FT-Raman spectral variation contained in the multi-modal spectral data using only 14 variables, which represents a 26.9% of increase for model interpretability. This increase may likely explain the slightly better characterization/grouping of the samples obtained by the VIP-refined HPLS model. An easy comparison of the clustering ability of the single-block and the multi-block PLS-based approaches is offered by **Fig. S16-S18** of **Supporting Information 1**, where some geological groups of specimens have been manually indicated in score plots for the PLS, the HPLS, and the VIP-HPLS cases; showing how the characterization/grouping of the geological specimens becomes more informative and precise when the multi-block HPLS model is used rather than a single-block PLS model, and afterwards, more detailed clusters (e.g., clusters discriminating types of silicates) are obtained when using the VIP method to generate a refined HPLS model.

***Table II.*** *Total number of model components (LV), percentage of explained X-variation ($R^2X$), and percentage of explained Y-variation ($R^2Y$) for the single-block PLS models and the multi-block HPLS models.*

| MODEL | LV | $R^2X$ | $R^2Y$ |
|---|---|---|---|
| ASD-NIR PLS | 11 | 99.4 | 52.5 |
| Micro-NIR PLS | 7 | 99.7 | 23.7 |
| Bruker-Raman PLS | 15 | 87.8 | 77.3 |
| i-Raman PLS | 7 | 97.4 | 43.1 |
| HPLS | 14 | 73.1 | 85.4 |
| VIP-refined HPLS | 14 | 100.0 | 68.7 |

## 4 Discussion

In this paper we investigated the potential use of multi-block chemometric methods to provide improved unsupervised characterization of compositionally complex materials through the integration of multi-modal spectrometric data using geological samples, with elemental composition determined by XRF, as an exemplar data set. The single-block modelling approach was based on partial least squares regression, the multiblock approach on hierarchical PLS modelling, and the multi-block dimensionality reduction for enhanced characterization/grouping on variable influence on projection feature selection.

### 4.1. PLS model performance for characterization of samples in single-spectral data

Partial least squares projections to latent structures [18,19] is commonly used for data interpretation, multivariate calibration, prediction of a response matrix **Y** from a descriptive matrix **X**, pattern recognition, and discriminant analysis. A PLS model decomposes the data matrices in their latent structures (latent variables, *a.k.a.* model components) that explain the different properties or sets of information (variance) of the data. Each latent variable (LV) is calculated as the product of the scores (t) and transposed loadings (p') for each model component *a*, i.e., $LV_a = t_a p_a'$. These latent variables are interpreted by generating multi-dimensional visualizations and inspecting their associated statistics. For our spectrometric data, this inspection led to the conclusion that each of the four spectroscopic techniques (ASD-NIR, Micro-NIR, Bruker-Raman and i-Raman) was able to characterize/group certain, but not all, geological samples types. In this paper, we have aimed to an unsupervised sample characterization/grouping, based on the fact that a PLS regression can be carried out with or without pre-defined classes (categories), as explained in the Introduction. PLS can be used for supervised classification, known as partial least squares discriminant analysis (PLS-DA), which requires the classes to be pre-defined; and then, these classes are used for separating the samples in a multi-dimensional space. On the other hand, PLS can be run without pre-defined classes (which in this paper is named "unsupervised" in terms of sample grouping), i.e., without any previous knowledge of the existent classes in the data; therefore, these PLS models perform a totally data-driven characterization/ grouping of the samples, which is preferred in fields such as geology or archaelogy. Is it also important to highlight that in order to have an adequate sample characterization in the hierarchical modelling context of this paper, several model validation strategies were followed [26]. In this paper, we used the PLS models also to assess the performance of the different spectroscopic techniques for examining and grouping geological samples. And, in addition, we found evidence that variable influence on projection (VIP) applied on PLS models highlights the most important regions of a spectrum for unsupervised characterization of geological samples.

### 4.2. Multi-block HPLS models for sample characterization in integrated multi-spectral data

We hypothesised that multi-block approaches would allow better discriminate of compositionally complex materials through the integration of multi-modal spectrometric data. For testing this in our multi-modal spectrometric data, a hierarchichal multi-block approach based on the PLS formalism was adopted. Hierarchical partial least squares (HPLS) [20] is a muti-block modelling technique that generates latent models of the original data matrices (sub-level modelling) and applies the PLS algorithm to the resulting scores (super-level modelling). In this paper, the generation of the HPLS models started by generating sub-level individual PLS models, all of them with the same samples, for each NIR or FT-Raman data set (block) using the XRF data as response block. The scores from these individual PLS models were then used as *super-variables* to form *super-matrices* that became the imputed **X** and **Y** data matrices in the HPLS model. The outputs of the HPLS model were *super-scores* and *super-loadings* able to explain the relations of the geological specimens and group them based on the information provided by both NIR and FT-Raman spectral data, rather than using only the information of one single type of spectroscopy. This yielded a holistic sample grouping, where the limitations of the individual NIR/Raman PLS models to cluster certain groups were overcome. Besides, it made the interpretation of the clusters easier and parsimonious providing a more complete characterization, yet not perfect, in one single model integrating NIR and Raman variances. This HPLS approach also allowed, by inspecting loadings and scores, an assessment of which spectroscopic techniques are more suitable for identifying and classifying different geological samples.

### 4.3. Variable influence on projection (VIP) for achieving improved sample characterization in multi-spectral HPLS models

Multi-block models based on the partial least squares algorithm (such as PLS, PLS-DA, $O_2$PLS, $O_n$PLS, etc.) often benefit from variable/feature selection methods, such as VIP [21,27] or MB-VIOP [22], to enhance their interpretability and/or classification power. To improve the sample characterization of our geological samples, the variable influence on projection (VIP) method was used to select the most important variables to explain and cluster the bedrock specimens. VIP is a variable selection method that works with both PLS and HPLS models since they have the same statistical principles. A feature-reduced second HPLS model was built using only the PLS scores (used as **X** super-variables of the HPLS model) that were assessed as important by the VIP algorithm (i.e., with VIP value > 1 *a.u.*). This second HPLS model was to evaluate whether the same (or better) characterization/grouping could be obtained with a multiblock model using only a reduced number of VIP selected PLS scores (i.e., using less **X** super-variables). We found that the HPLS model built with the VIP-selected PLS scores (i.e., with lower dimensionality) achieved a better unsupervised sample characterization than the original HPLS model, without any disadvantage when compared to the original model. Due to have lower dimensions, the VIP-refined HPLS model is computationally more efficient than the original HPLS model, with more easily interpretable loading plots thanks to the reduced number of features.

### 4.4. Multi-disciplinar applicability of multi-block HPLS and VIP-HPLS models for sample characterization

When comparing the single-block and the multi-block strategies, the HPLS models, where both NIR and FT-Raman variances complement each other, provided a more complete clustering of the geological specimens. Multi-block modelling showed advantages for building more parsimonious models, which made them easier to interpret; as well as more computationally efficient, potentially allowing the adoption of this approach to multi-modal spectrometric data sets across numerous scientific disciplines. In addition to the component-wise dimensionality reduction achieved by the extraction of the latent structures contained in our compositionally complex multi-modal data, a post-modelling VIP feature-wise dimensionality reduction was carried out, reducing the roughly 3700 original measured wavelengths and wavenumbers (variables used in the four single-block PLS models) to 14 input variables (PLS scores of the sub-level individual models) used in the VIP-simplified HPLS model. Therefore, simplicity and efficiency were achieved without any loss of information.

Spectroscopic techniques have some limitations due to sample incompatability with the instrument, size, shape and colour, or the analytical approach. Some of these challenges can be overcome by using other non-destructive techniques, such as hyperspectral imaging [28,29], with the adoption of chemometric methods, based on projections and latent structures [26,30]. However, hyperspectral instruments are less available and so there remains merit in combining different types of spectroscopy from more widely used instruments (e.g, NIR, FT-Raman, and XRF) with multi-block chemometric methods as an affordable and efficient alternative.

In conclusion, here we analysed five multi-modal spectrometric data sets (originated by two NIR, two FT-Raman, and one XRF instruments) with PLS and HPLS models and found that the combination of NIR and FT-Raman spectra, using XRF data as response, in HPLS multi-block models resulted in a good unsupervised (i.e., in this paper's context, without any need to pre-define sample classes or categories) characterization/grouping of the samples. This PLS multi-modal chemometric approach had the advantage of providing a single, parsimonious, and more efficient multi-block model (HPLS), rather than several (and not so efficient) single-block models (PLS). The second main finding confirmed that reducing the multi-block HPLS model dimensions by applying the variable influence on projection (VIP) method improved the model interpretation, as well as the characterization of certain groups of samples, such as the silicates. We would like to emphasise that the methodology evaluated in this paper applied to geological and archaelogical data, is also compatible with any type of multivariate data across a broad range of fields such as environmetal chemistry, physics, -omics, medicine or artificial intelligence, where integration of multi-modal data presents the opportunity to improve sample characterization/grouping (as within this study), or to help uncover causal relationships between environmental source/chemicals and biological responses.

## Acknowledgments

The authors would like to thank the Kempe Foundation (grant no. SMK-1749) for financial support, the Department of Ecology and Environmental Science (Umeå University, Sweden) for the use of the geological collection, and Mats Eriksson from the Environmental Archaeology Lab (Umeå University, Sweden) for assisting in the spectra acquisition.

## Author contributions

All authors (JL, PG, ISM, BGP) contributed to the study conception, design and conceptualization. Data acquisition was done by JL, the pictures of the instrumentation (Fig. 1 and Fig. S1) were taken by PG, and all figures/visualizations of the paper were created by BGP. Methodological strategy was planned by PG, BGP, JL and ISM. Data analysis, including data preprocessing and modelling, was performed by BGP and PG. The Supporting Information was produced by BGP and PG. The manuscript was written by BGP; and ISM, PG and JL commented and reviewed it. All authors read and approved the final manuscript.

## Declaration of competing interests

The authors declare no competing interests.

## Supporting Information

Two sets of **Supporting Information (1 and 2)** are included in the online version of this article at the publisher's web site. **Supporting Information 1** is available as pdf file and contains **Tables S1-S8** and **Figures S1-S18**. **Supporting Information 2** is a repository that contains additional statistics and visualizations available upon request from authors.

## Data availability

The data will be available in the infrastructure databases SEAD and SWEDIGARK.

## References

1. Martin FL, German MJ, Wit E, Fearn T, Ragavan N, Pollock HM. Identifying variables responsible for clustering in discriminant analysis of data from infrared microspectroscopy of a biological sample. *Journal of Computational Biology*. 2007;**14**(9):1176-1184. doi:10.1089/CMB.2007.0057.

2. Kumar K, Sivabalan S, Ganesan S, Mishra AK. Discrimination of oral submucous fibrosis (OSF) affected oral tissues from healthy oral tissues using multivariate analysis of in vivo fluorescence spectroscopic data: A simple and fast procedure for OSF diagnosis. *Analytical Methods*. 2013;**5**:3482-3489. doi:10.1039/c3ay40352a.

3. Hamany Djande CY, Piater LA, Steenkamp PA, Tugizimana F, Dubery IA. A metabolomics approach and chemometric tools for differentiation of barley cultivars and biomarker discovery. *Metabolites*. 2021;**11**(9):578. doi:10.3390/METABO11090578/S1.

4. Zhang S, Li X, Geng T, et al. Using machine learning to predict soil lead relative bioavailability. *J Hazard Mater*. 2025;**483**:136515. doi:10.1016/J.JHAZMAT.2024.136515.

5. Mishra P, Roger J-M, Jouan-Rimbaud-Bouveresse D, et al. Recent trends in multi-block data analysis in chemometrics for multi-source data integration. 2021. doi:10.1016/j.trac.2021.116206.

6. Brereton RG. Pattern recognition in chemometrics. *Chemometrics and Intelligent Laboratory Systems*. 2015;**149**:90-96. doi:10.1016/J.CHEMOLAB.2015.06.012.

7. Jenkins R. *X-Ray Fluorescence Spectrometry*. Wiley; 1999. doi:10.1002/9781118521014.

8. Timothy Oyedotun TD. X-ray fluorescence (XRF) in the investigation of the composition of earth materials: a review and an overview. *Geology, Ecology, and Landscapes*. 2018;**2**(2):148-154. doi:10.1080/24749508.2018.1452459.

9. Flude S, Haschke M, Storey M. Application of benchtop micro-XRF to geological materials. *Mineral Mag*. 2017;**81**(4):923-948. doi:10.1180/MINMAG.2016.080.150.

10. Guo F, Clemens S, Liu X, et al. Application of XRF Scanning to Different Geological Archives. *Earth and Space Science*. 2021;**8**(9):e2020EA001589. doi:10.1029/2020EA001589.

11. Burns DA, Ciurczak EW. *Handbook of Near-Infrared Analysis*. CRC Press; 2007. doi:10.1201/9781420007374.

12. Gardiner DJ, Graves PR. *Practical Raman Spectroscopy*. Springer Berlin Heidelberg; 1989. doi:10.1007/978-3-642-74040-4.

13. Edwards HGM, Hutchinson IB, Ingley R, Parnell J, Vítek P, Jehlička J. Raman Spectroscopic Analysis of Geological and Biogeological Specimens of Relevance to the ExoMars Mission. *Astrobiology*. 2013;**13**(6):543-549. doi:10.1089/AST.2012.0872.

14. Andò S, Garzanti E. Raman spectroscopy in heavy-mineral studies. *Geol Soc Spec Publ*. 2014;**386**(1):395-412. doi:10.1144/SP386.2.

15. Shankar V. Field Characterization by Near Infrared (NIR) Mineral Identifiers- A New Prospecting Approach. *Procedia Earth and Planetary Science*. 2015;**11**:198-203. doi:10.1016/J.PROEPS.2015.06.025.

16. Badura I, Dąbski M. Reflectance spectroscopy in geology and soil sciences: Literature review. *Quaestiones Geographicae*. 2022;**41**(3):157-167. doi:10.2478/QUAGEO-2022-0031.

17. Wold S, Esbensen K, Geladi P. Principal component analysis. *Chemometrics and Intelligent Laboratory Systems*. 1987;**2**(1-3):37-52. doi:10.1016/0169-7439(87)80084-9.

18. Wold S, Martens H, Wold H. The multivariate calibration-problem in chemistry solved by the PLS method. *Lecture Notes in Mathematics*. 1983;**973**:286-293.

19. Geladi P, Kowalski BR. Partial least-squares regression: a tutorial. *Anal Chim Acta*. 1986;**185**:1-17.

20. Wold S, Hellberg S, Lundstedt T, Sjöström M, Wold H. PLS modeling with latent variables in two or more dimensions. In: *Symposium on PLS Model Building: Theory and Application.* . Frankfurt am Main; 1987.

21. Galindo-Prieto B, Eriksson L, Trygg J. Variable influence on projection (VIP) for orthogonal projections to latent structures (OPLS). *J Chemom*. 2014;**28**(8):623-632. doi:10.1002/cem.2627.

22. Galindo-Prieto B, Geladi P, Trygg J. Multiblock variable influence on orthogonal projections (MB-VIOP) for enhanced interpretation of total, global, local and unique variations in OnPLS models. *BMC Bioinformatics*. 2021;**22**(1):1-27. doi:10.1186/s12859-021-04015-9.

23. Sjölander M, Linderholm J, Geladi P, Buckland PI. Quartzite complexities: Non-destructive analysis of bifacial points from Västerbotten, Sweden. 2024. doi:10.1016/j.jasrep.2024.104381.

24. Barnes RJ, Dhanoa MS, Lister SJ. Standard Normal Variate Transformation and De-trending of Near-Infrared Diffuse Reflectance Spectra. *Appl Spectrosc*. 1989;**43**(5):772-777. https://doi.org/10.1366/0003702894202201.

25. Geladi P, MacDougall D, Martens H. Linearization and Scatter-Correction for Near-Infrared Reflectance Spectra of Meat. *Appl Spectrosc*. 1985;**39**(3):491-500. https://doi.org/10.1366/0003702854248656.

26. Galindo-Prieto B, Westad F. Classification in hyperspectral images by independent component analysis, segmented cross-validation and uncertainty estimates. *Journal of Spectral Imaging*. 2018;**7**. doi:10.1255/jsi.2018.a4.

27. Galindo-Prieto B, Trygg J, Geladi P. A new approach for variable influence on projection (VIP) in O2PLS models. *Chemometrics and Intelligent Laboratory Systems*. 2017;**160**:110-124. doi:10.1016/j.chemolab.2016.11.005.

28. Sciuto C, Geladi P, La Rosa L, Linderholm J, Thyrel M. Hyperspectral Imaging for Characterization of Lithic Raw Materials: The Case of a Mesolithic Dwelling in Northern Sweden. *Lithic Technology*. 2019;**44**(1):22-35. doi:10.1080/01977261.2018.1543105.

29. Amigo JM, Babamoradi H, Elcoroaristizabal S. Hyperspectral image analysis. A tutorial. *Anal Chim Acta*. 2015;**896**:34-51. doi:10.1016/j.aca.2015.09.030.

30. Dumarey M, Galindo-Prieto B, Fransson M, Josefson M, Trygg J. OPLS methods for the analysis of hyperspectral images—comparison with MCR-ALS. *J Chemom*. 2014;**28**(8):687-696. doi:10.1002/cem.2628.

31. Panigrahi P, Sharma RK, Hasan M, Parihar SS. Deficit irrigation scheduling and yield prediction of 'Kinnow' mandarin (Citrus reticulate Blanco) in a semiarid region. *Agric Water Manag*. 2014;**140**:48-60. doi:10.1016/J.AGWAT.2014.03.018.

32. O’Brien NA, Hulse CA, Friedrich DM, et al. Miniature near-infrared (NIR) spectrometer engine for handheld applications. *https://doi.org/101117/12917983*. 2012;**8374**:31-38. doi:10.1117/12.917983.

33. Beć KB, Grabska J, Huck CW. Miniaturized NIR Spectroscopy in Food Analysis and Quality Control: Promises, Challenges, and Perspectives. *Foods 2022, Vol 11, Page 1465*. 2022;**11**(10):1465. doi:10.3390/FOODS11101465.

**Supporting information 1 of:**

***Multi-block chemometric approaches to the unsupervised spectral characterization of geological samples***

Beatriz Galindo-Prieto[1,2,*], Ian S. Mudway[1,2], Johan Linderholm[3,*], Paul Geladi[4]

[1] Environmental Research Group, School of Public Health, Faculty of Medicine, Imperial College London, London, UK

[2] MRC Centre for Environment and Health, School of Public Health, Faculty of Medicine, Imperial College London, London, UK

[3] Environmental Archaeology Laboratory, Dept. of Historical, Philosophical and Religious Studies, Umeå University, Umeå, Sweden

[4] Biomass Technology and Chemistry, Swedish University of Agricultural Sciences, Umeå, Sweden

[*] Corresponding authors (b.galindo-prieto@imperial.ac.uk, johan.linderholm@umu.se)

ORCID's:

Beatriz Galindo-Prieto: https://orcid.org/0000-0001-8776-8626

Ian S. Mudway: https://orcid.org/0000-0003-1239-5014

Johan Linderholm: https://orcid.org/0000-0001-7471-8195

Paul Geladi: https://orcid.org/0000-0001-6618-5931

# FIGURES

**Fig. S1** Probes of the (a) ASD-NIR, (b) Micro-NIR, (c) i-Raman, and (d) XRF instruments. The measurements of the diameters of the probes are shown in the measuring tape (units: mm):

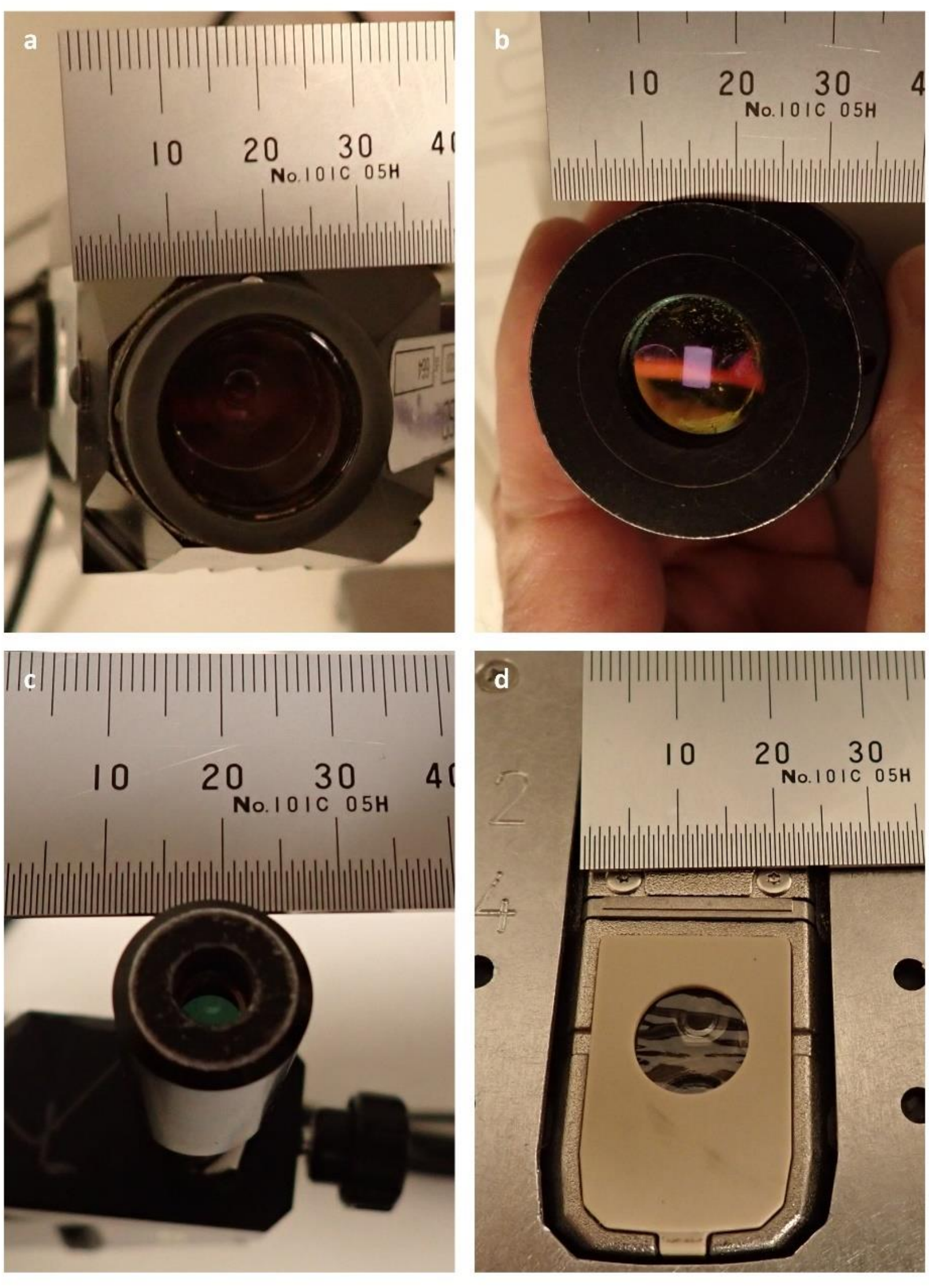

**Fig. S2** Raw data and pre-processed data from the ASD-NIR instrument:

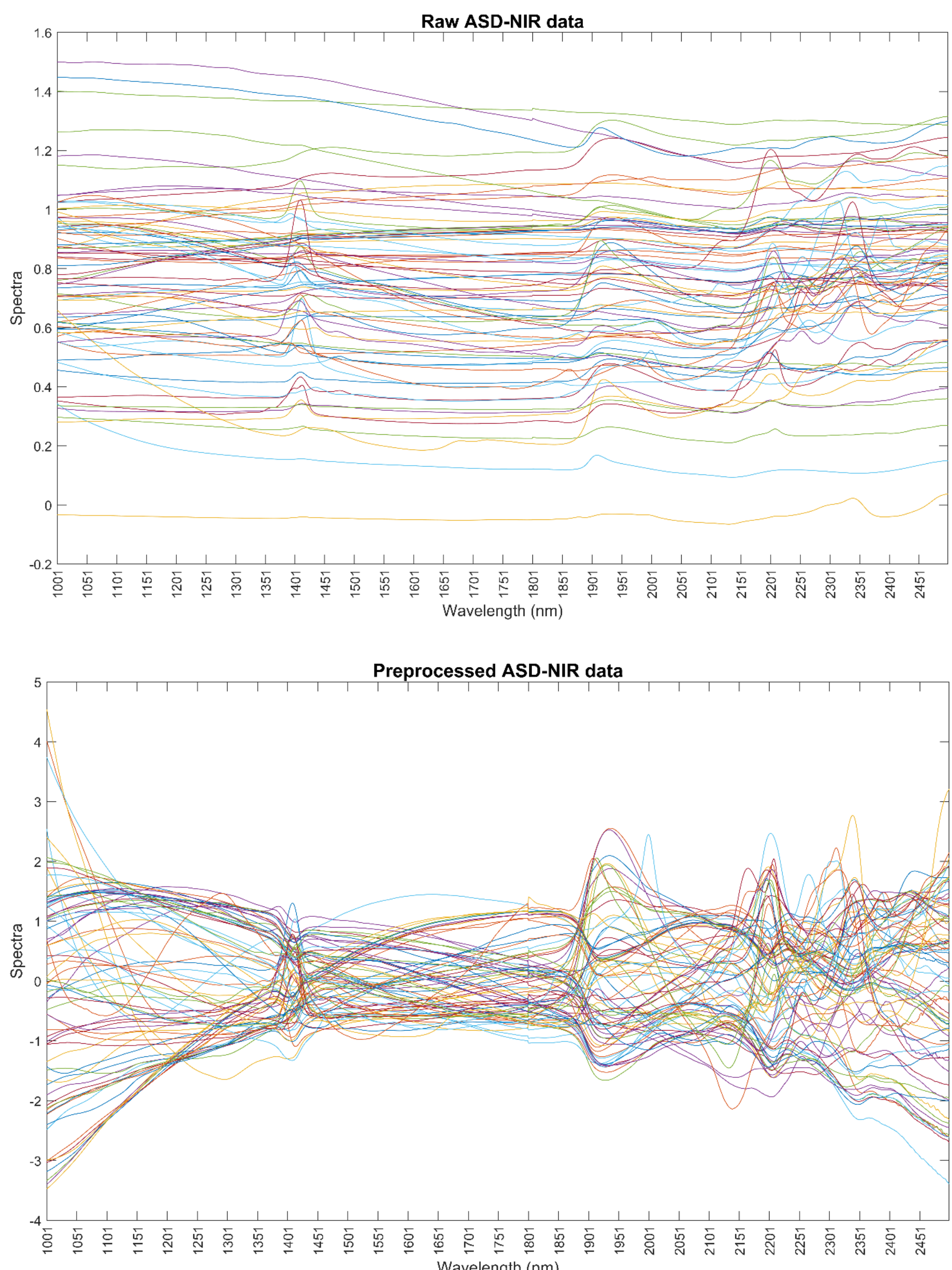

**Fig. S3** Raw data and pre-processed data from the Micro-NIR instrument:

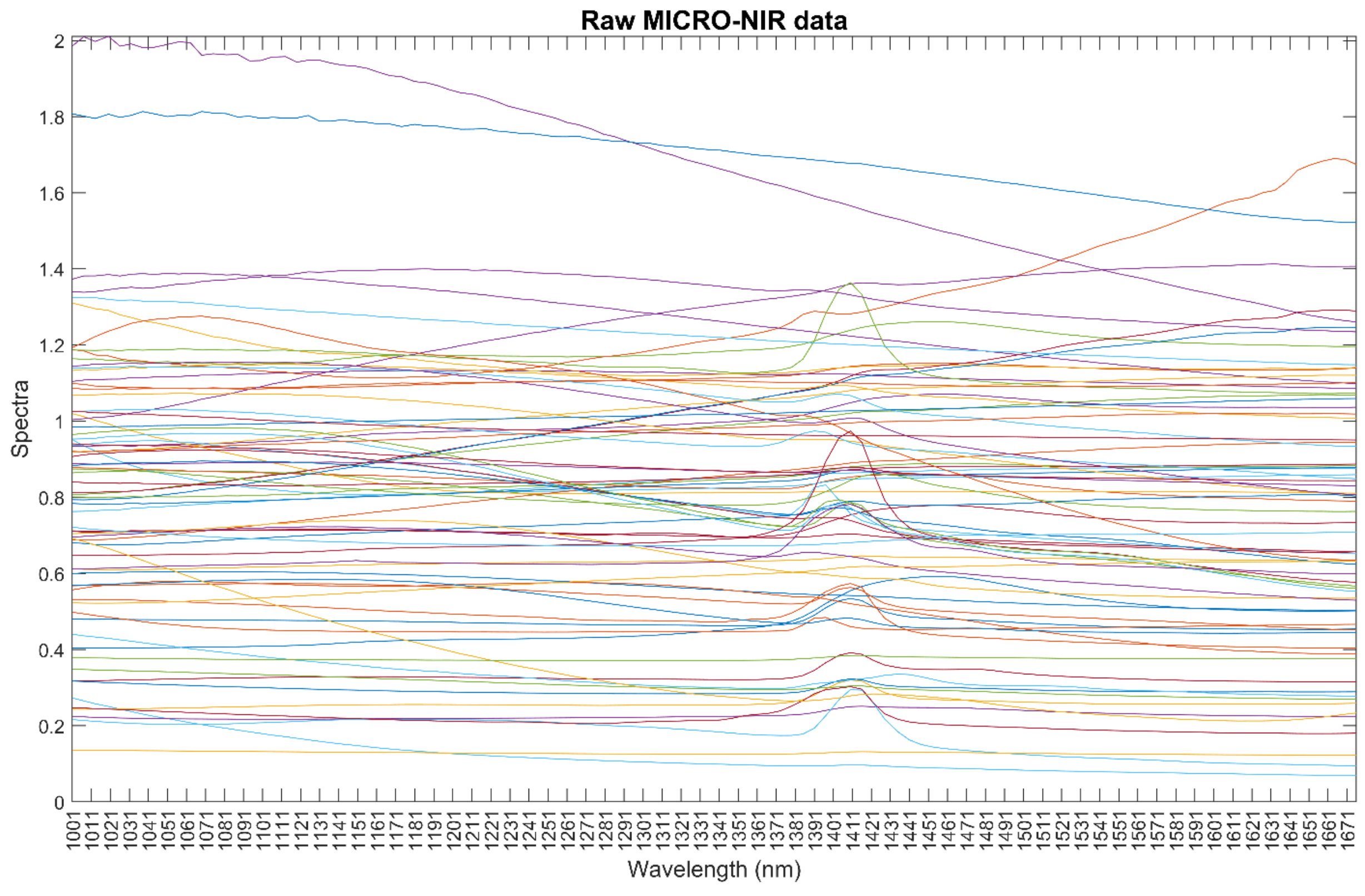

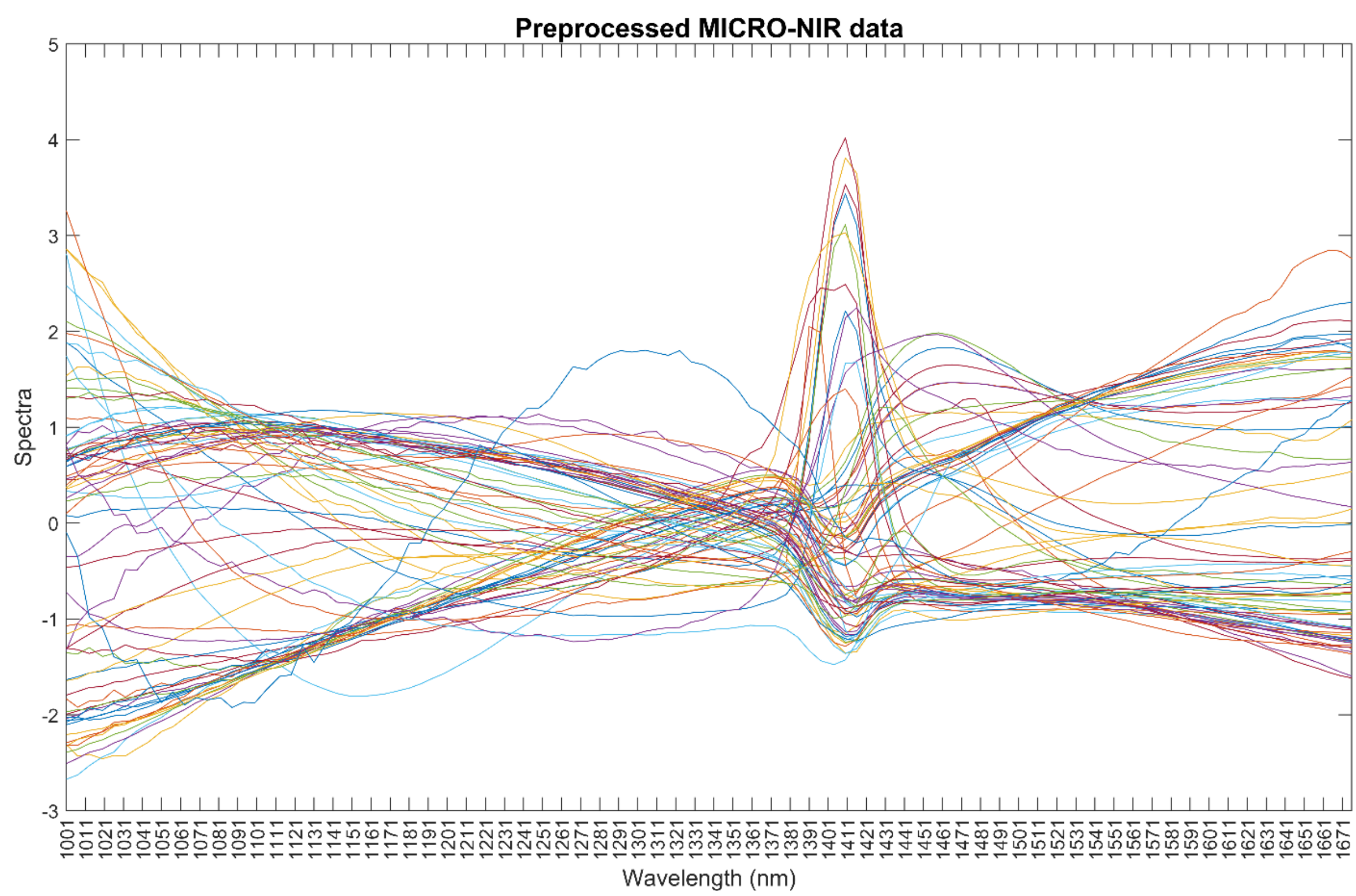

**Fig. S4** Raw data and pre-processed data from the Bruker-Raman instrument:

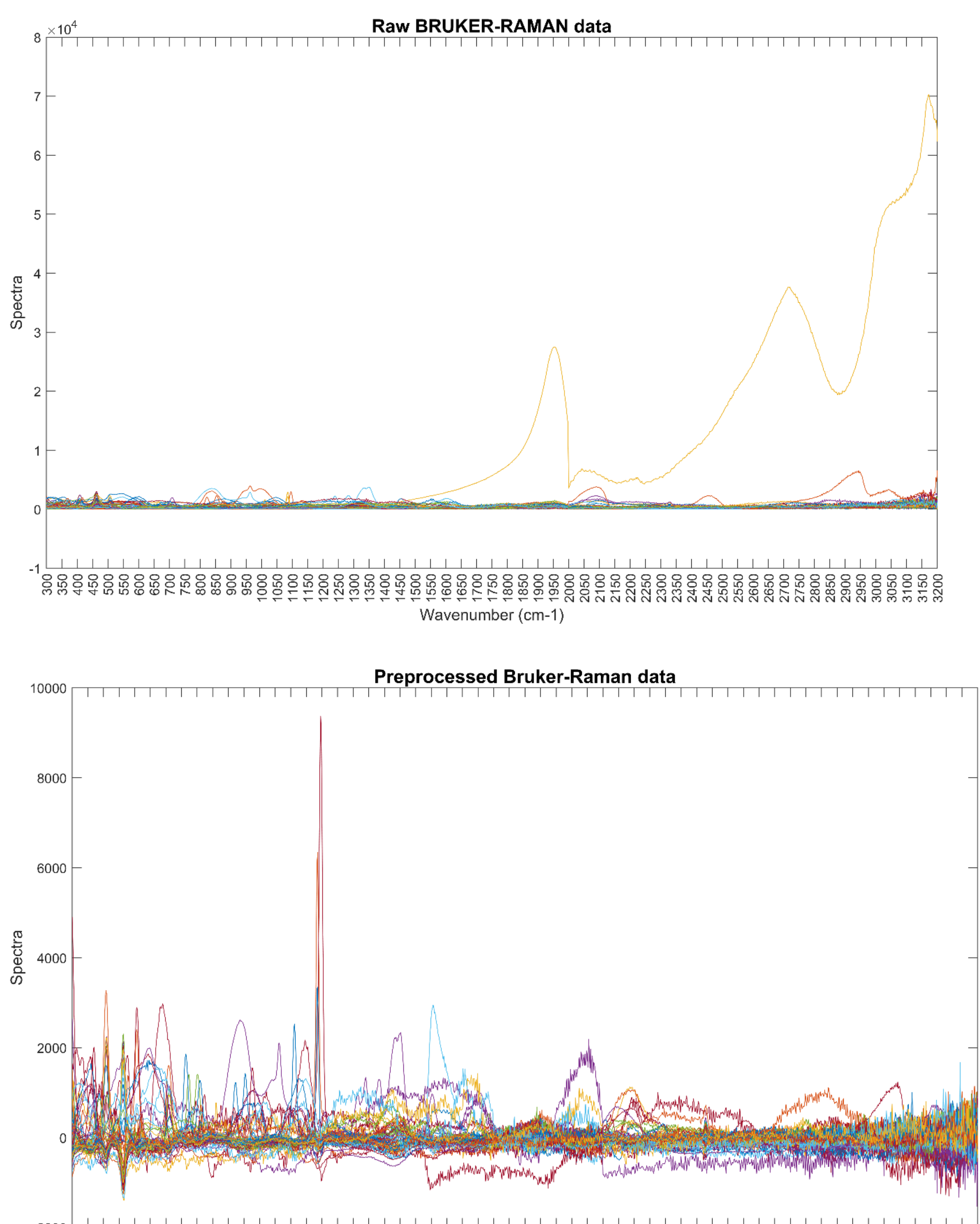

**Fig. S5** Raw data and pre-processed data from the i-Raman instrument:

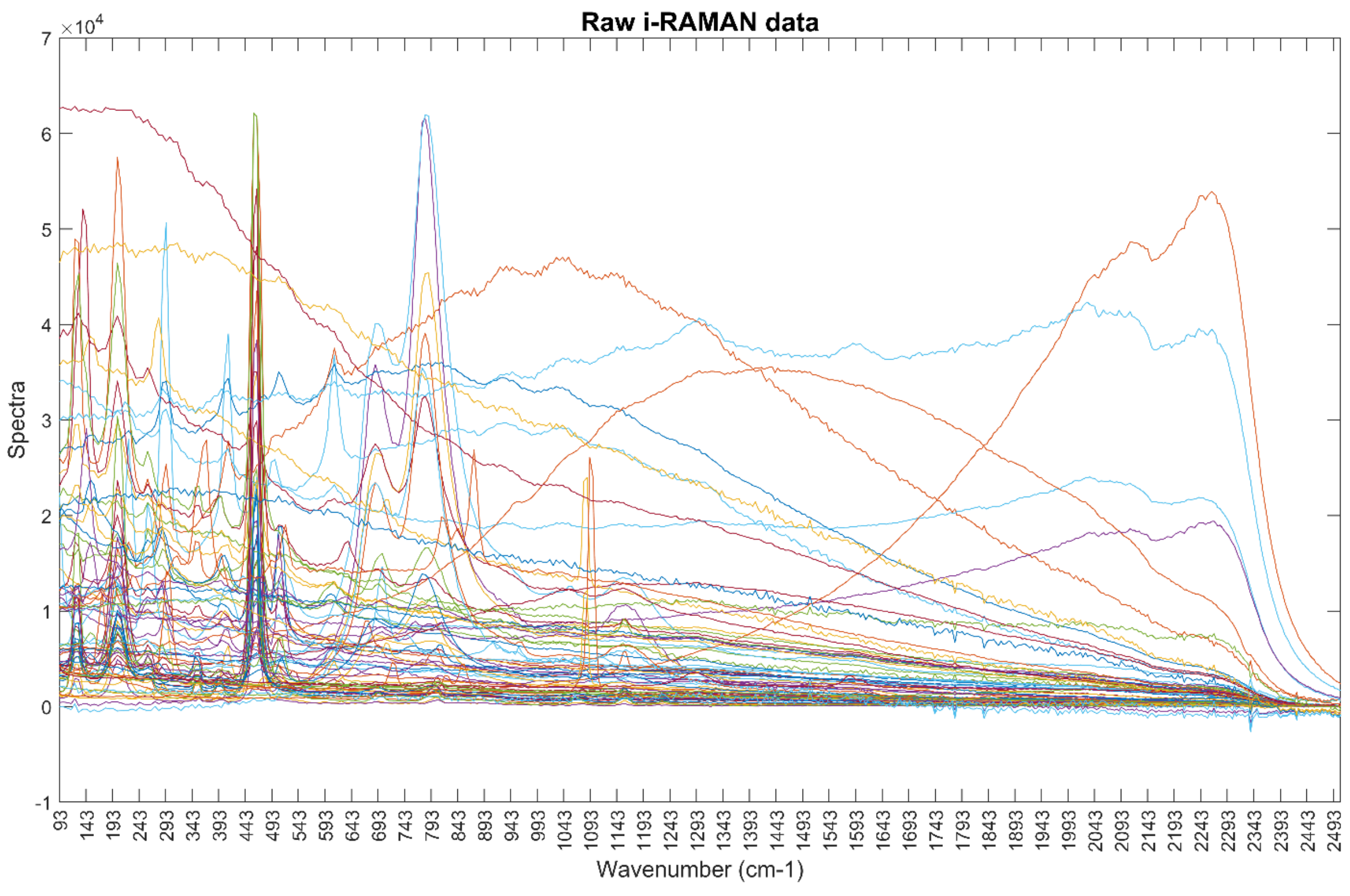

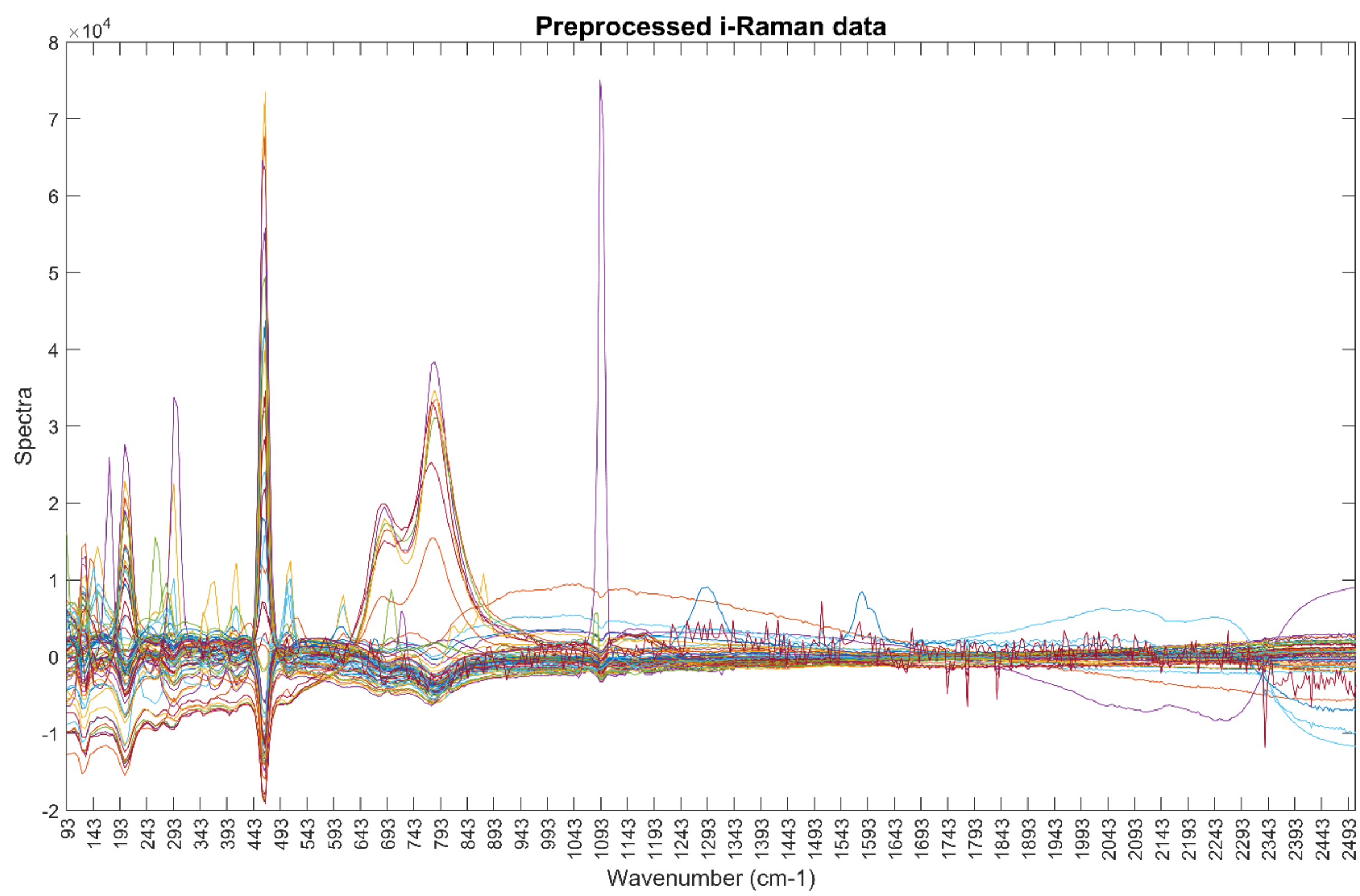

**Fig. S6** Raw data and pre-processed data from the XRF instrument:

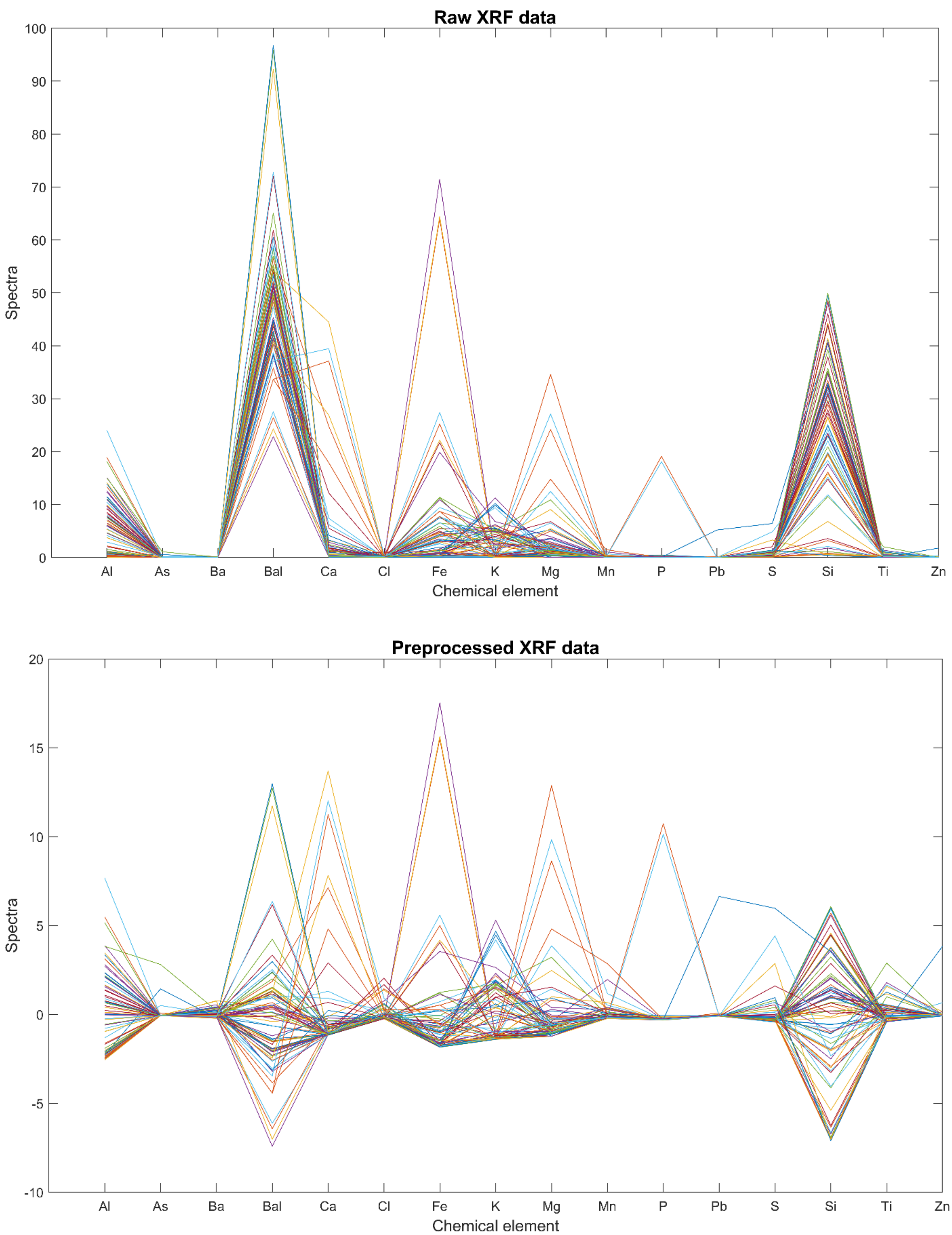

**Fig. S7** Loading plots for LV1 and LV2 of the four single-block PLS models as (**a**) scatter plots and (**b**) line plots:

**(a)** Scatter plot

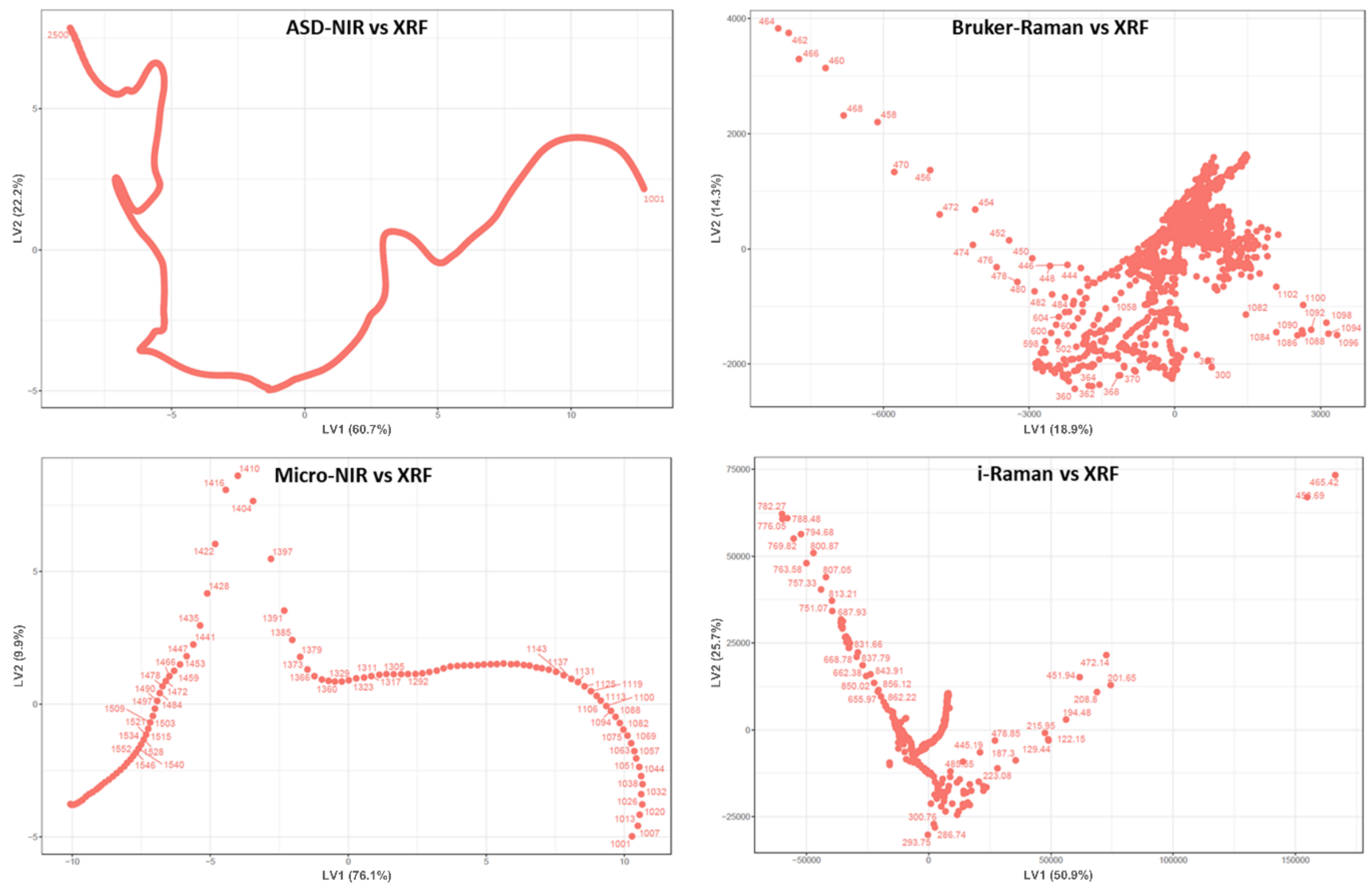

**(b)** Line plots

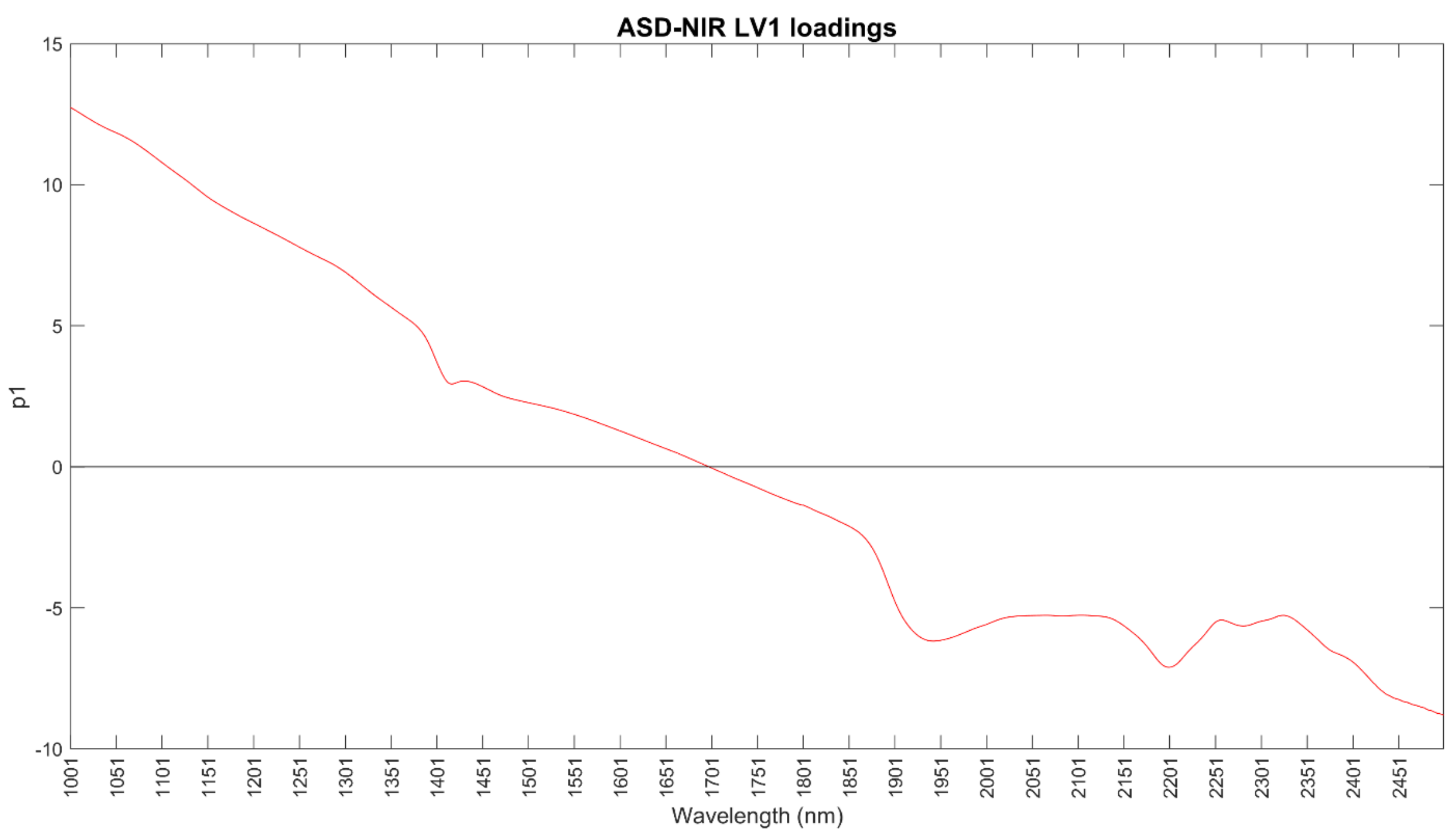

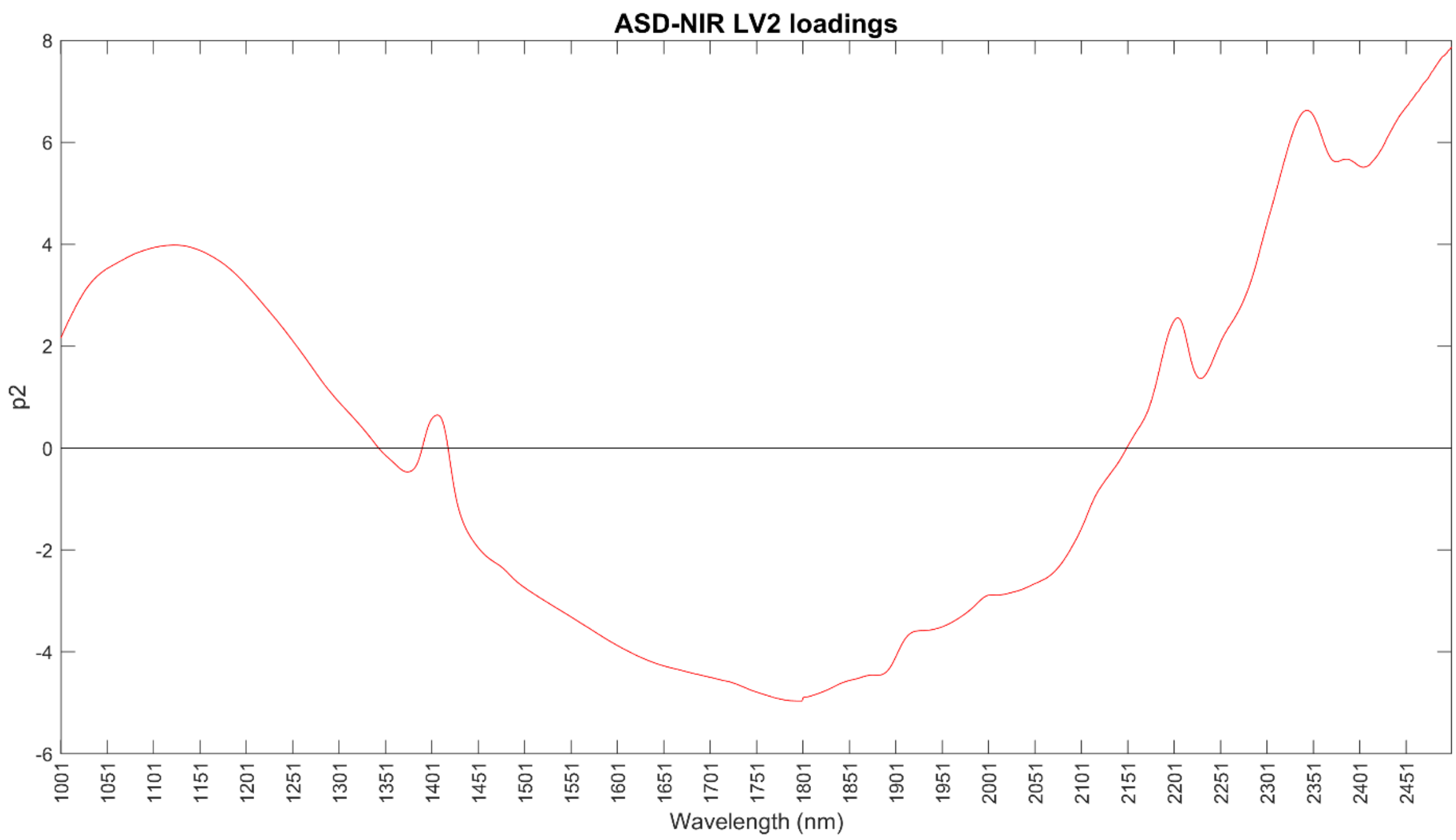

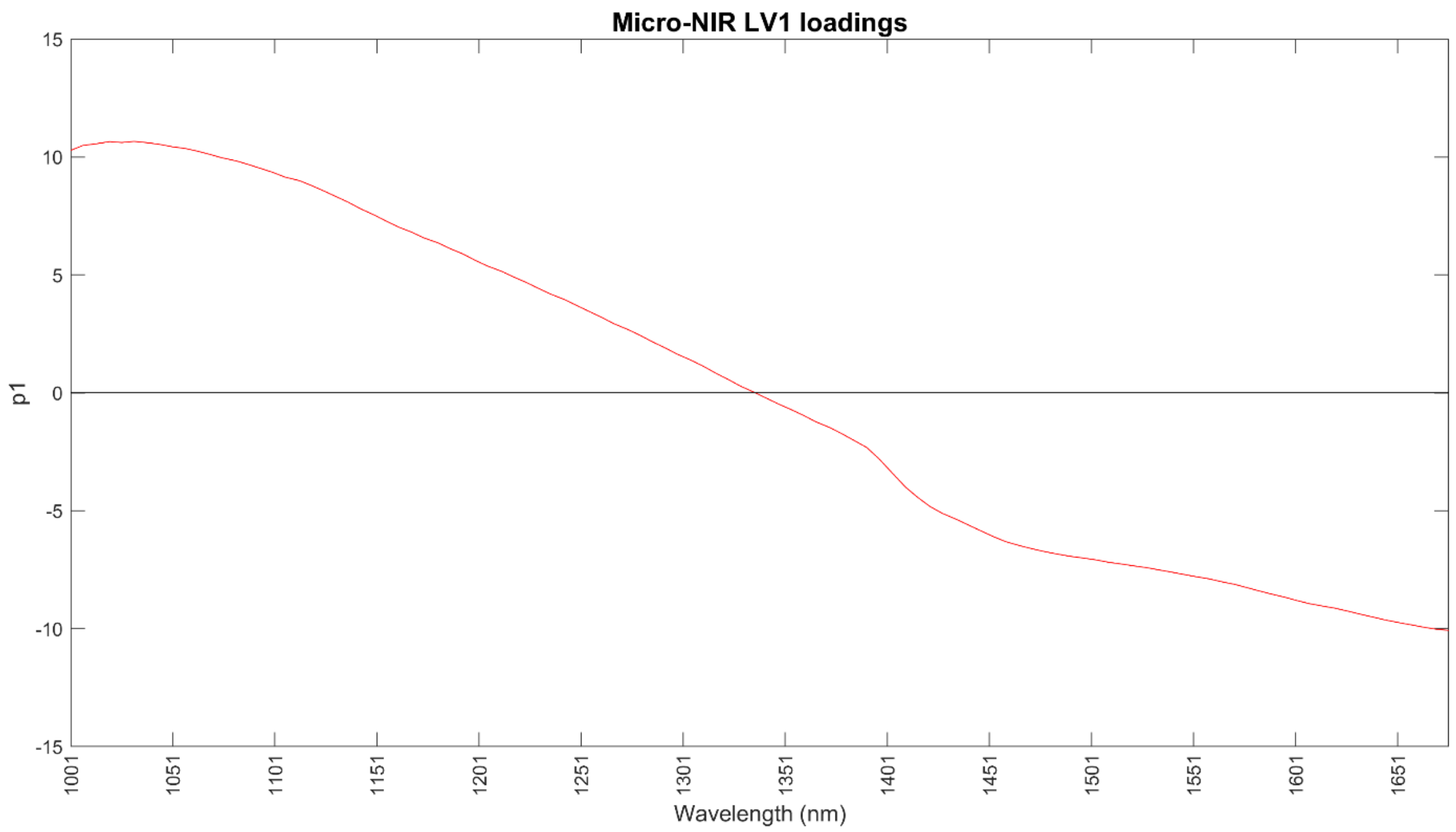

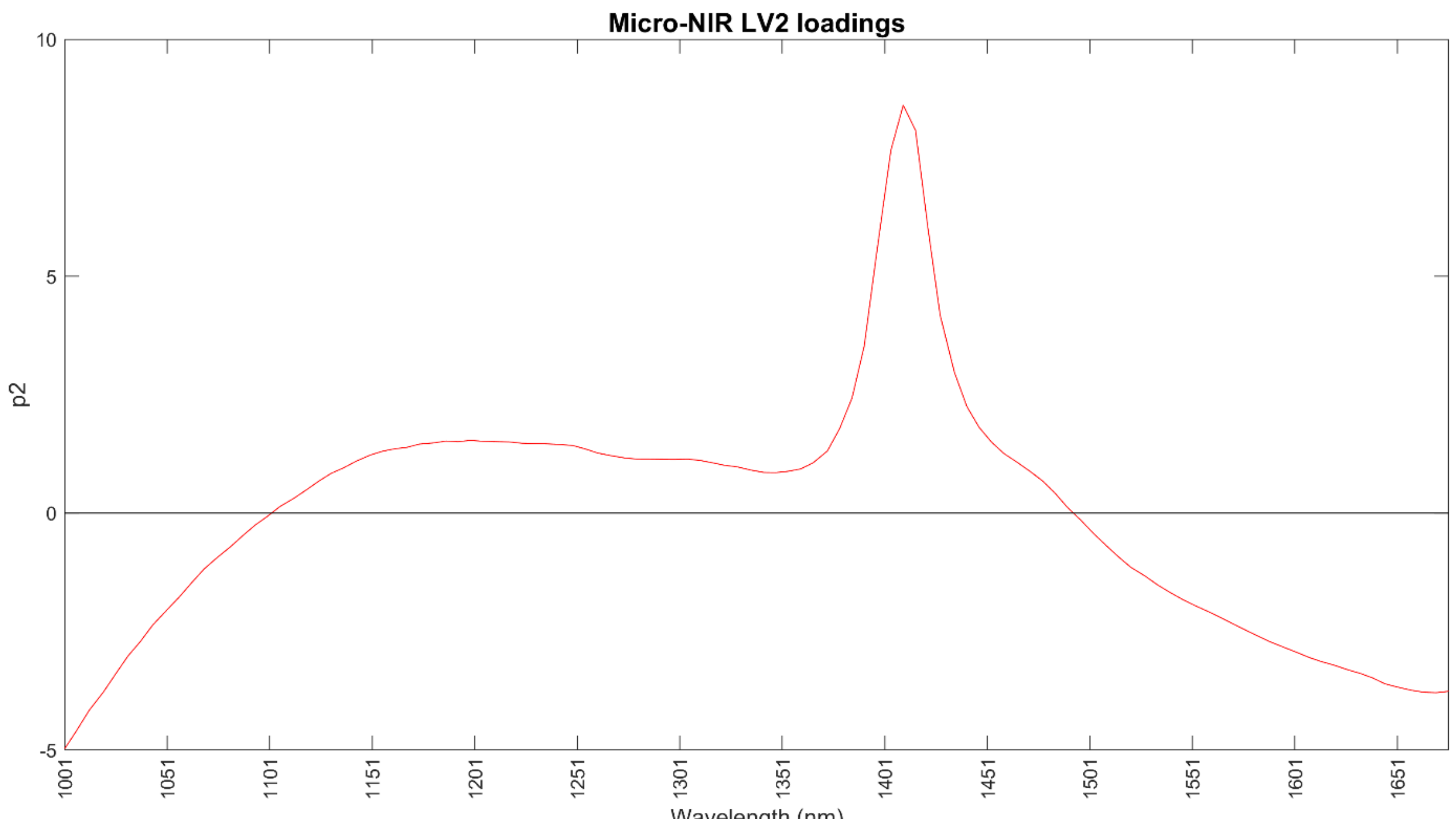

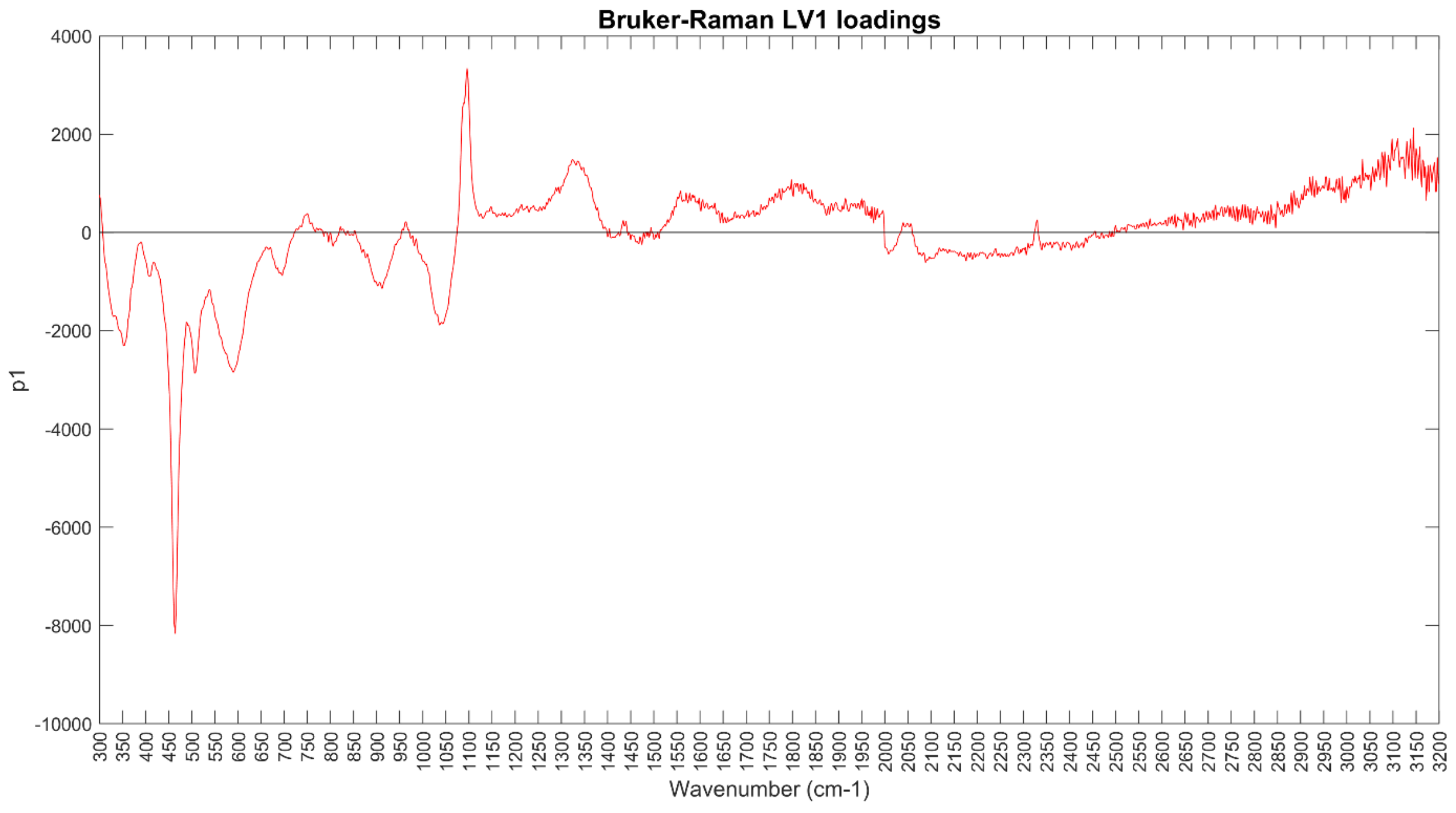

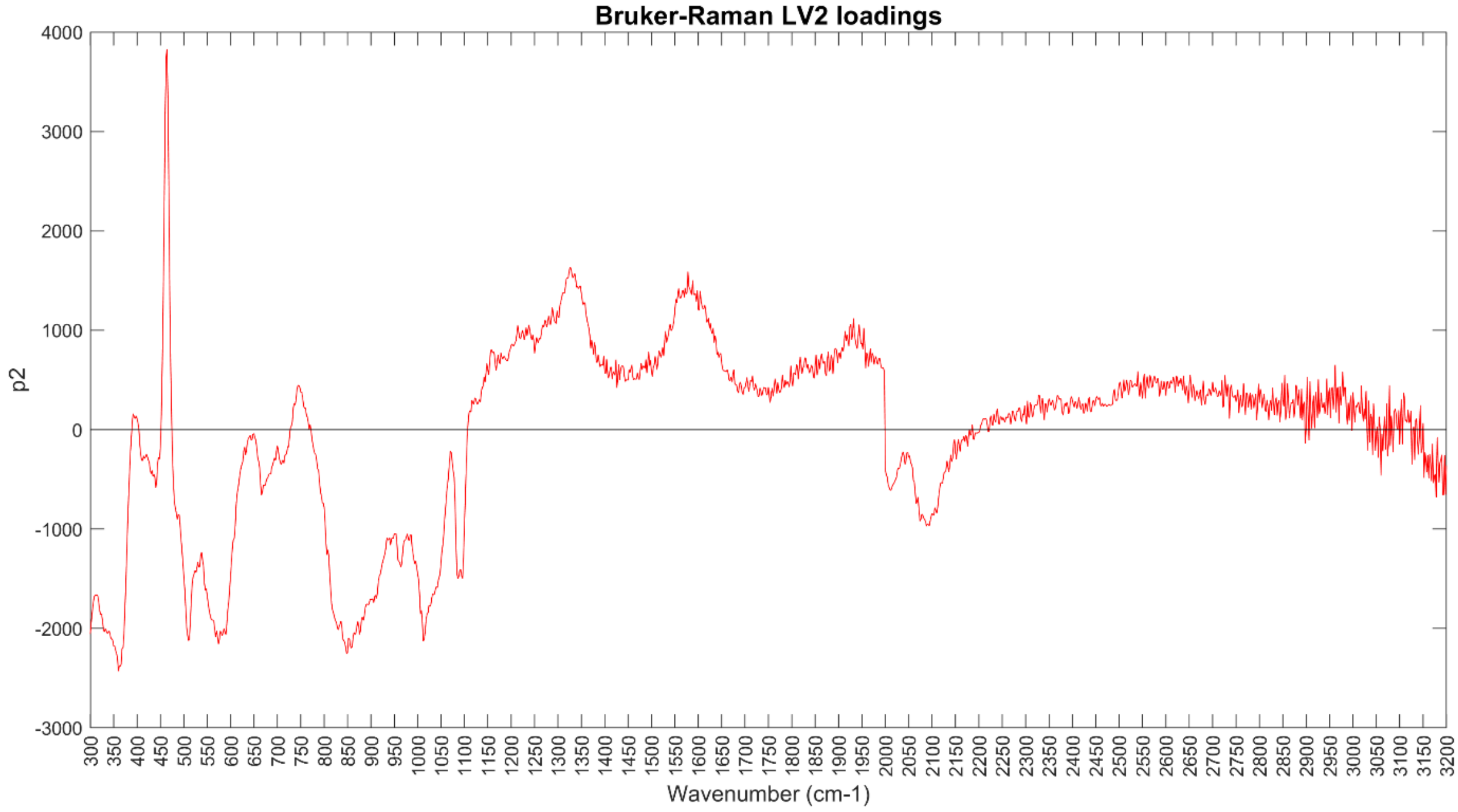

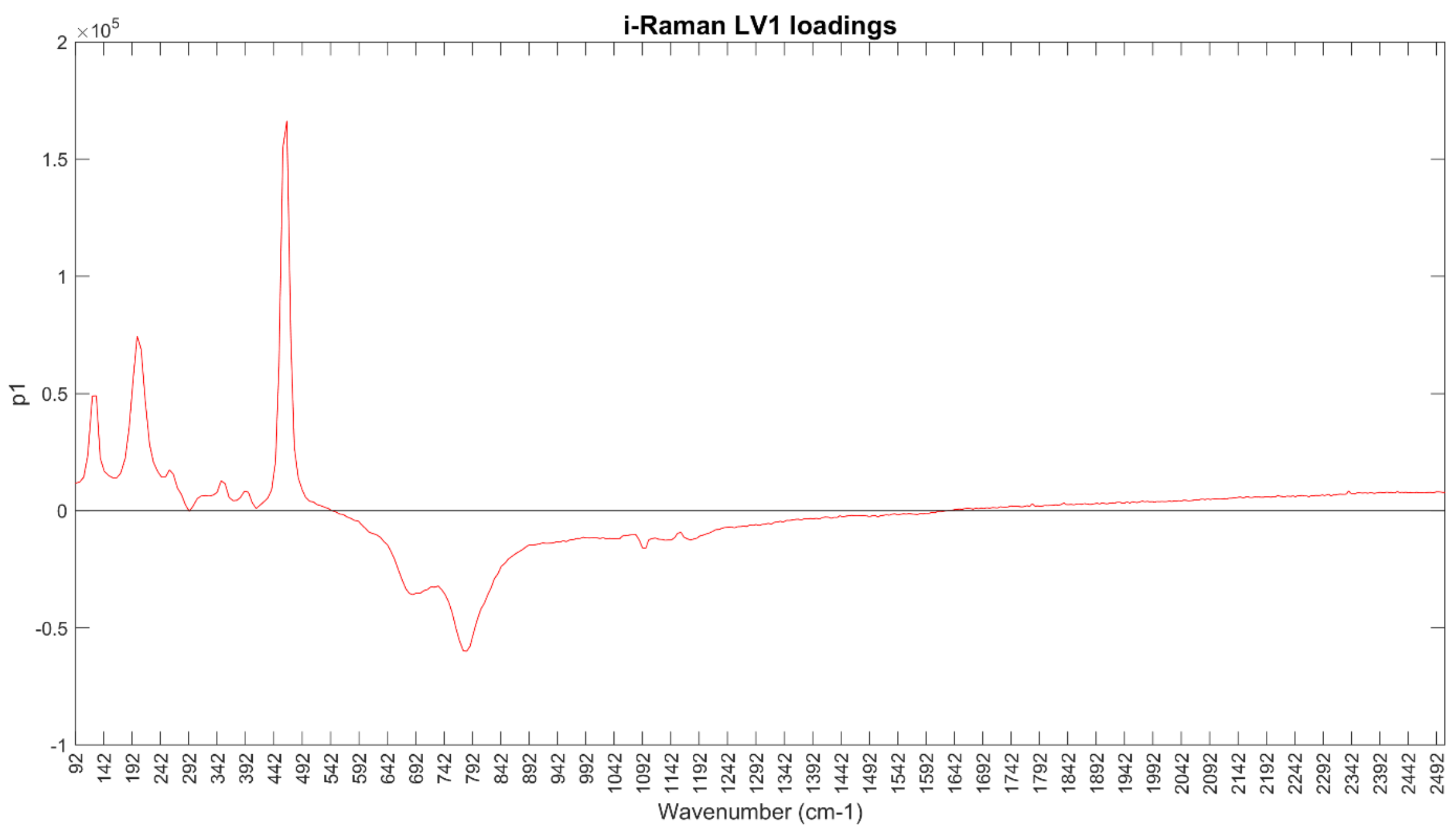

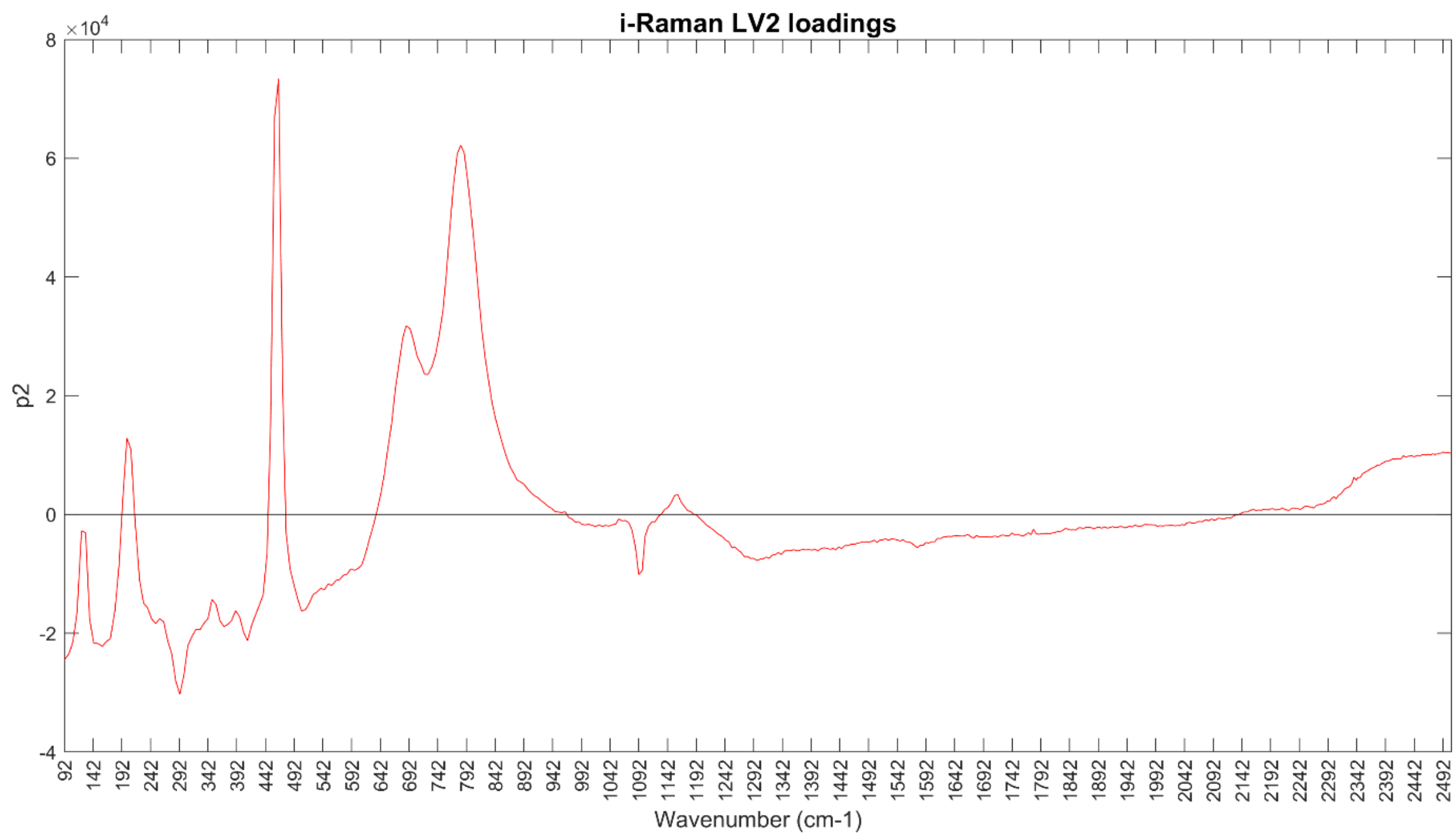

**Fig. S8** Elemental composition of metamorphic carbon samples:

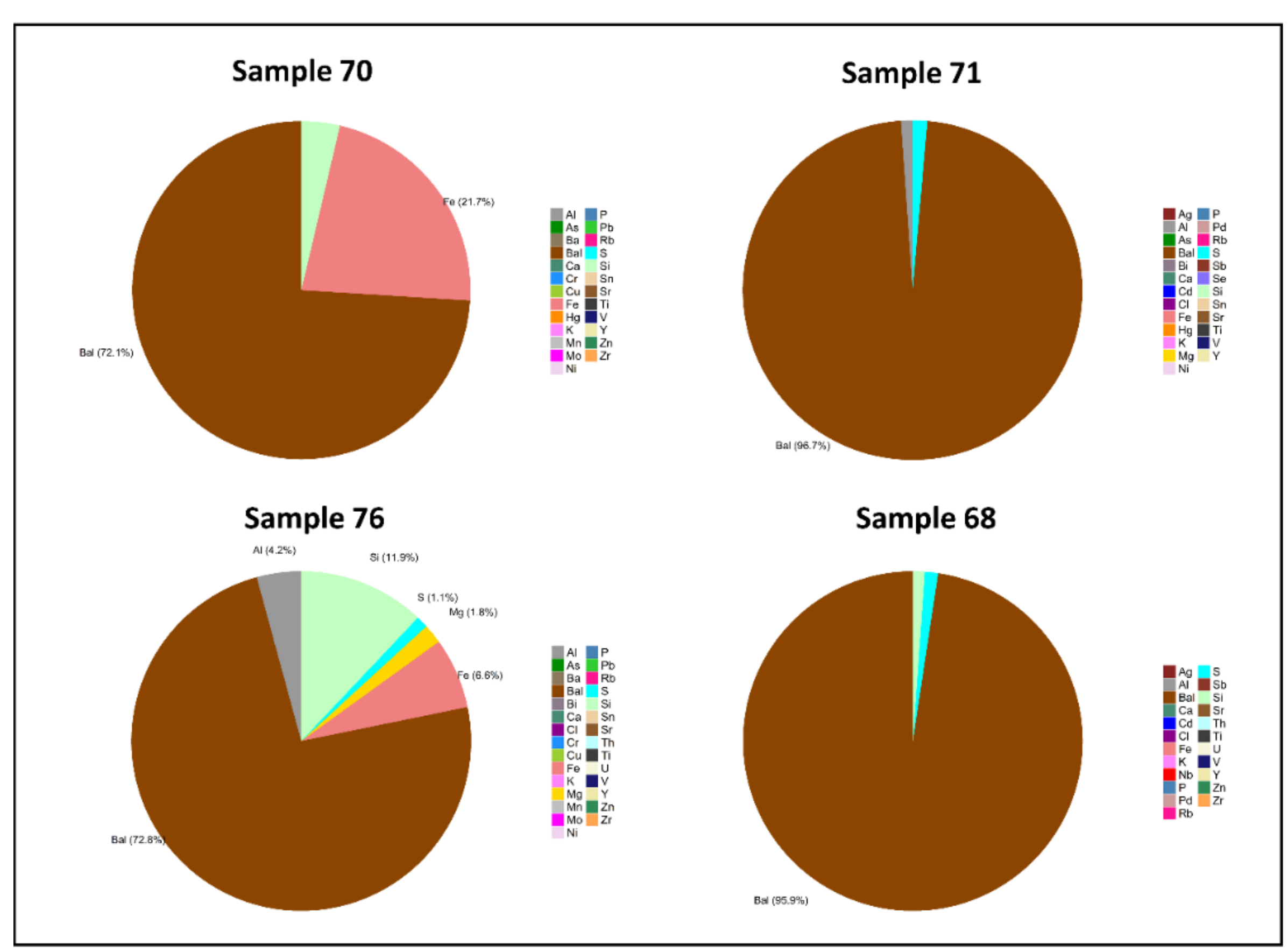

**Fig. S9** VIP for the Bruker-Raman PLS model using XRF data as response:

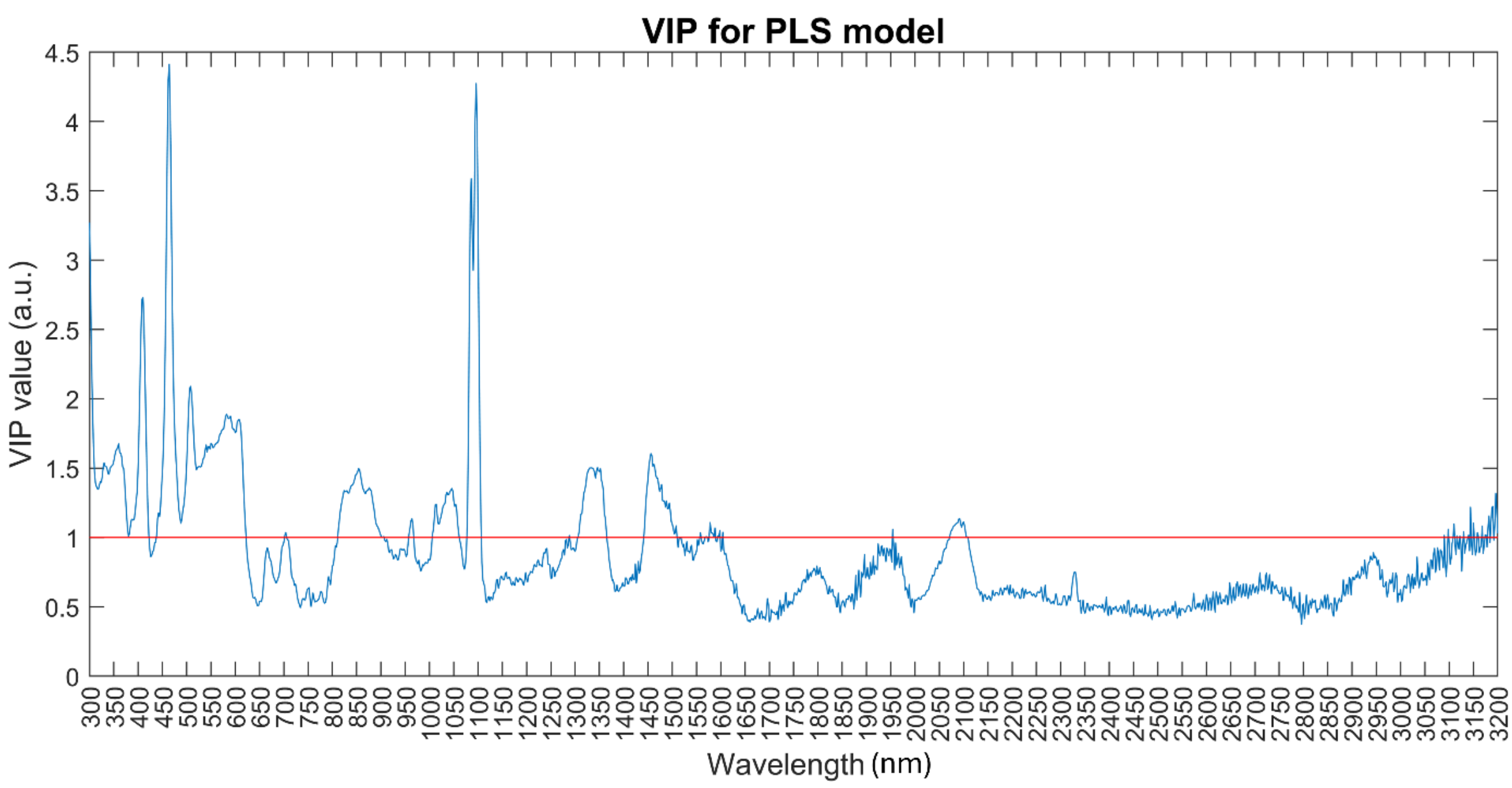

**Fig. S10** Biplot of the Bruker-Raman PLS model using only some of the VIP most relevant wavenumber variables:

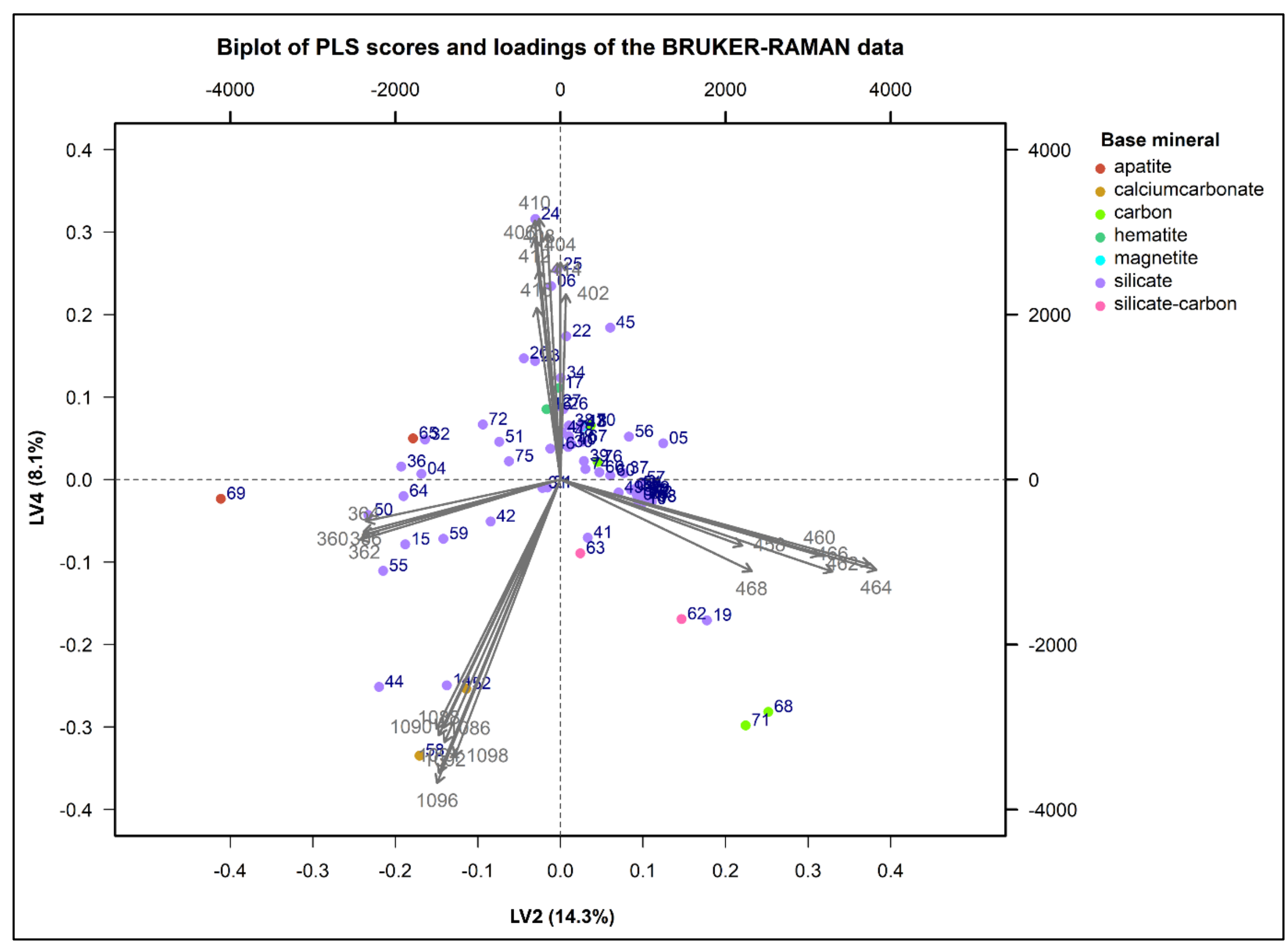

**Fig. S11** Sample grouping plots for an HPLS model built from single-block PCA models (LV1 and LV2):

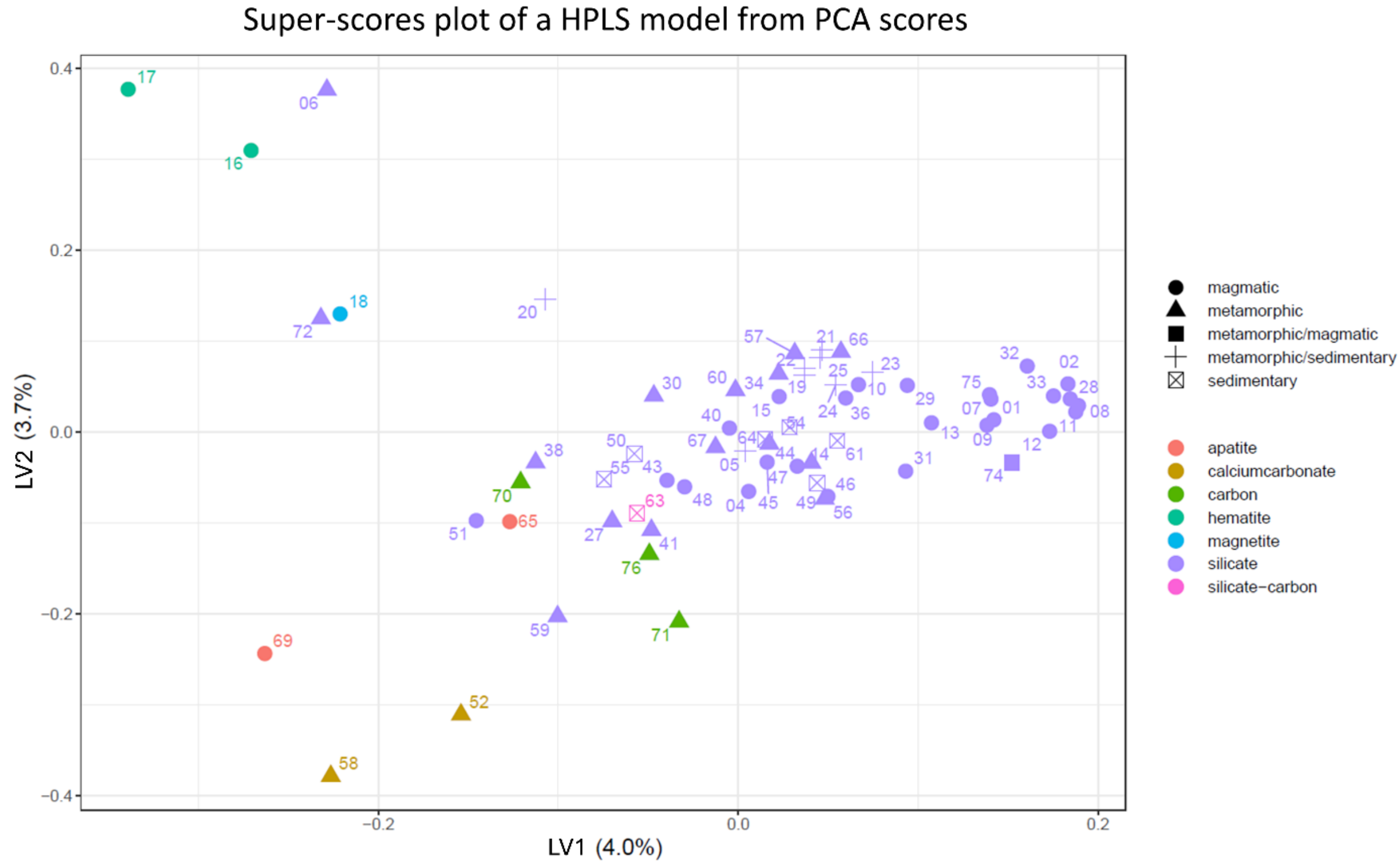

**Fig. S12** Super-loadings plots for LV1 and LV2 of (**a**) the HPLS and (**b**) the VIP-HPLS multi-block models:

**(a)**

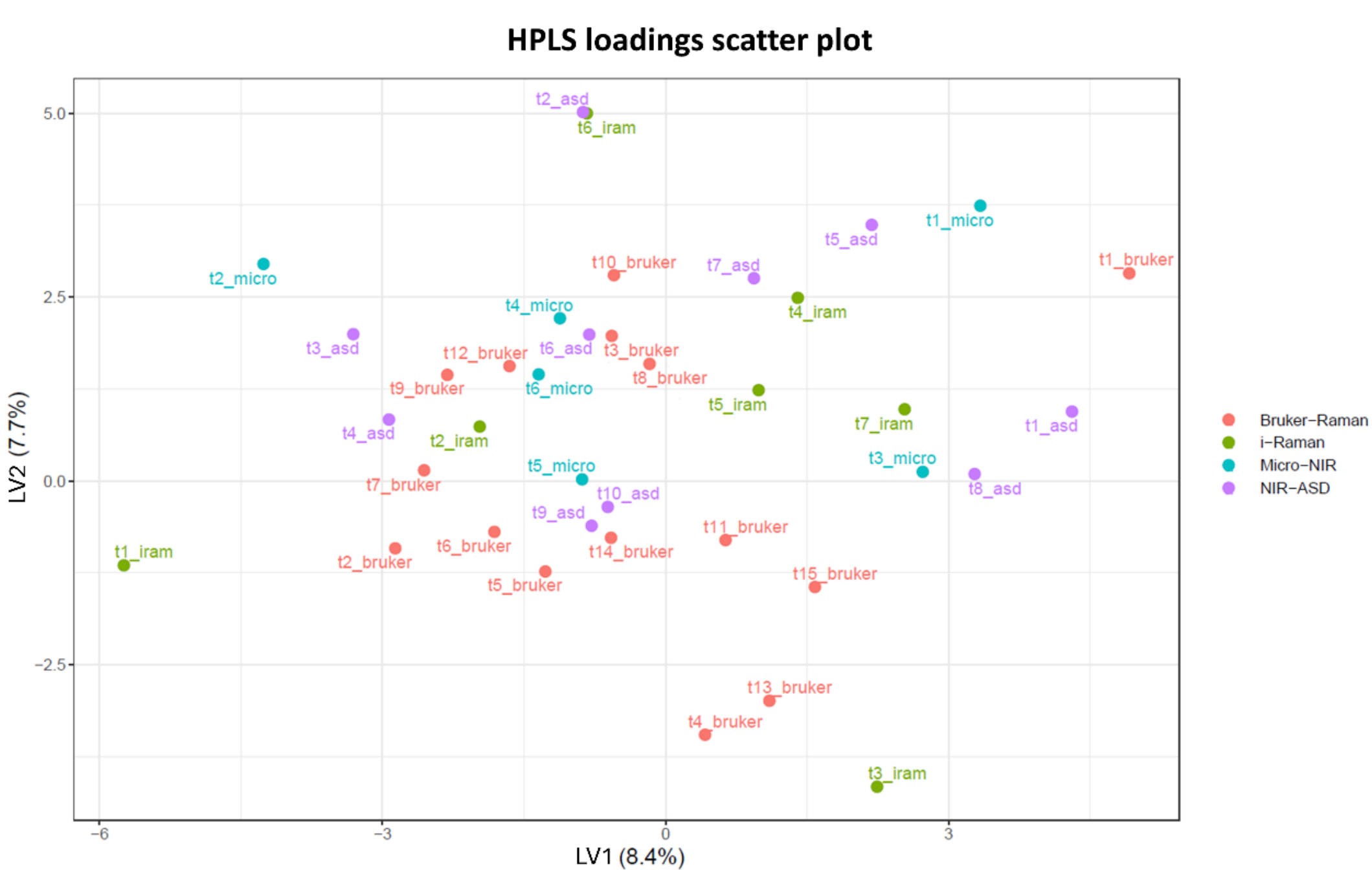

**(b)**

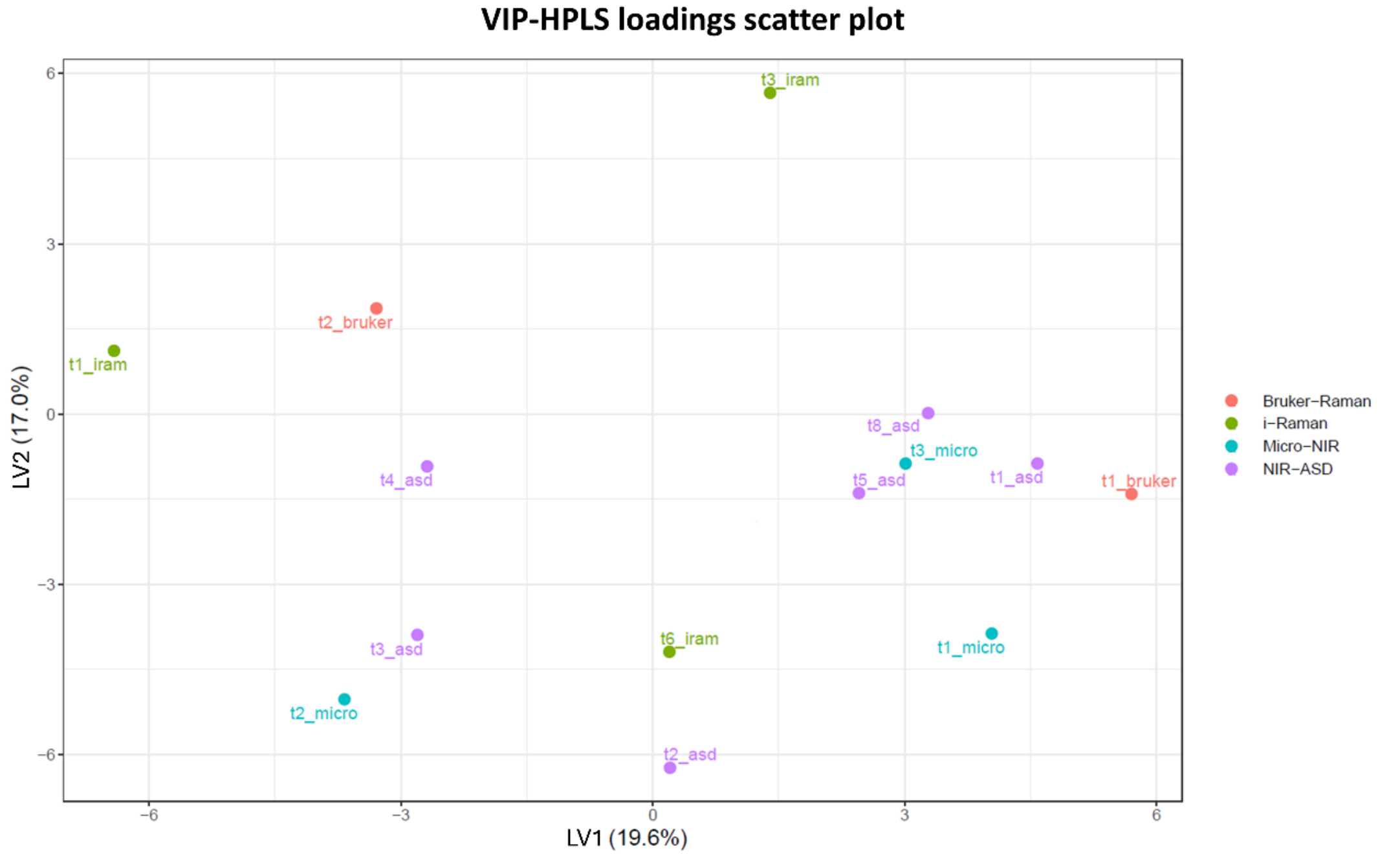

**Fig. S13** Elemental composition of the hematite samples:

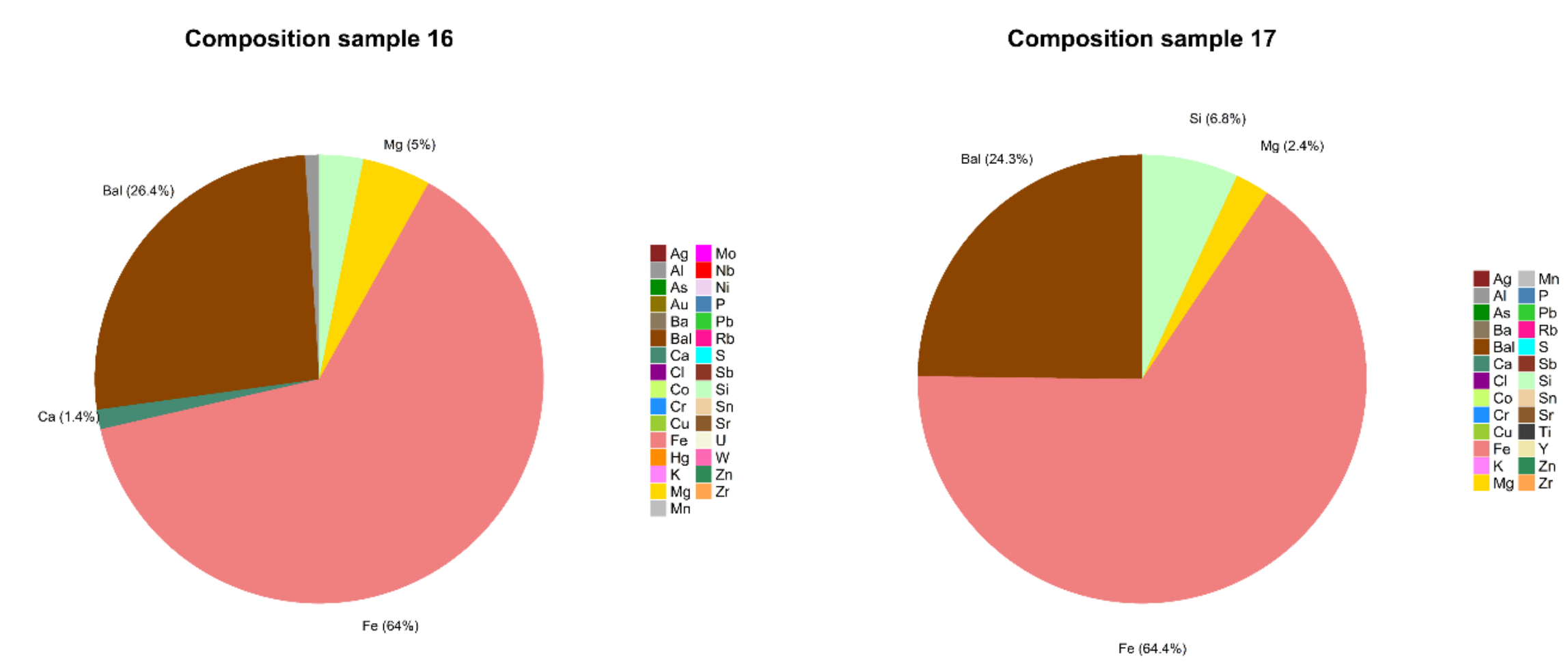

**Fig. S14** Elemental composition of the apatite samples:

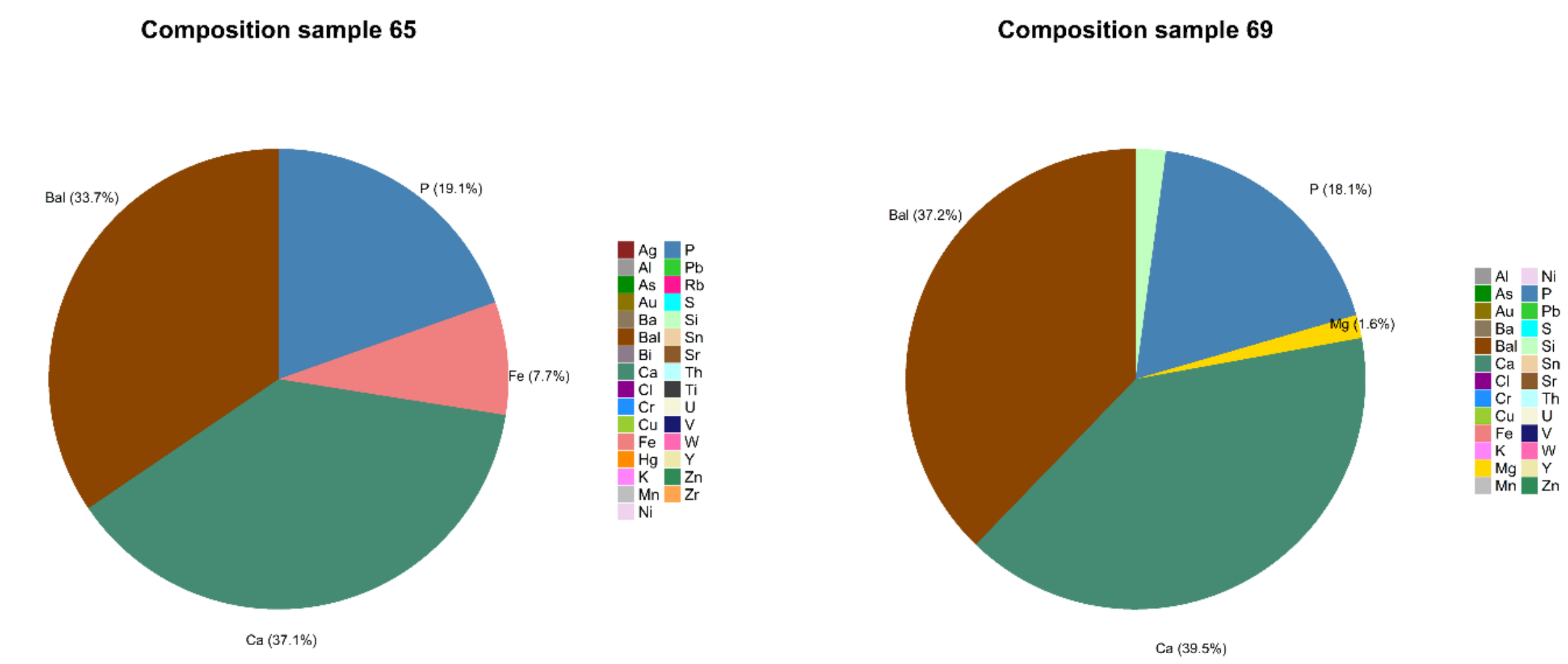

**Fig. S15** Elemental composition of the calcium-carbonate samples:

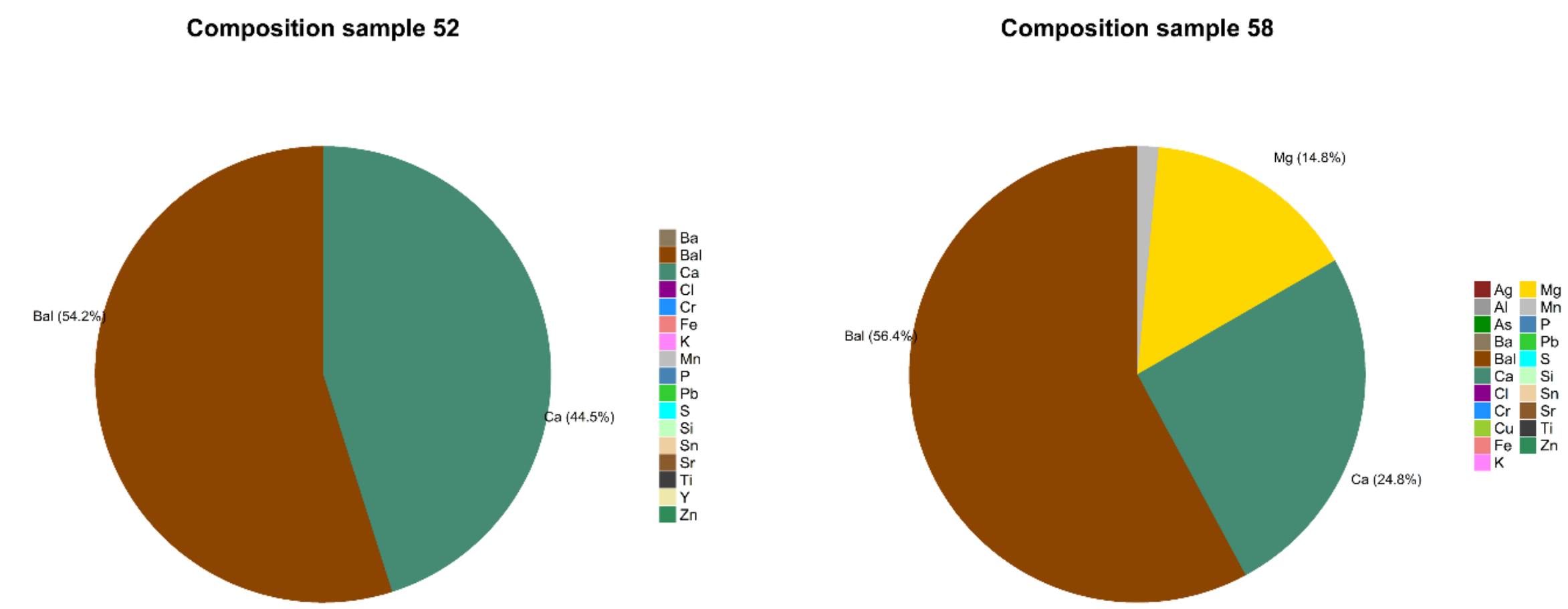

**Fig. S16** Scores plot for LV1 and LV2 of the FT-Raman PLS models with clusters marked according to mineral information for the i-Raman case:

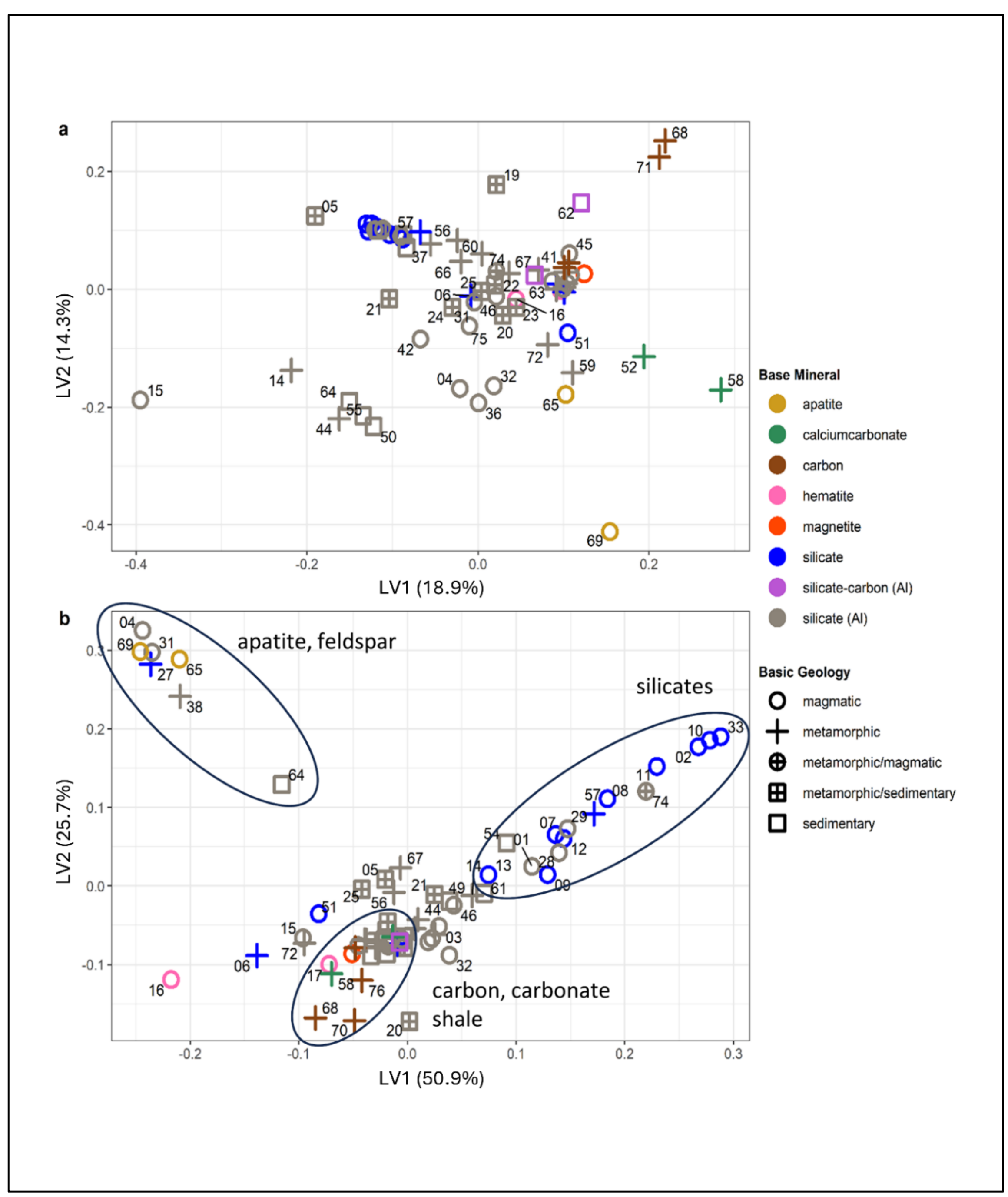

PLS scores for LV1 and LV2 for the (**a**) Bruker-Raman and (**b**) i-Raman models. Clusters are indicated by black ellipses, and cluster labelling corresponds to mineral information. The fact that LV1 is stronger in the i-Raman case relates to the sampling as the field of view is small and when the laser beam hits a mineral the spectra will be very distinct (for that specific mineral) compared to the Bruker Raman.

**Fig. S17** Scores plot of LV1 and LV2 for the HPLS model of all five spectral data sets with clusters marked by mineral information:

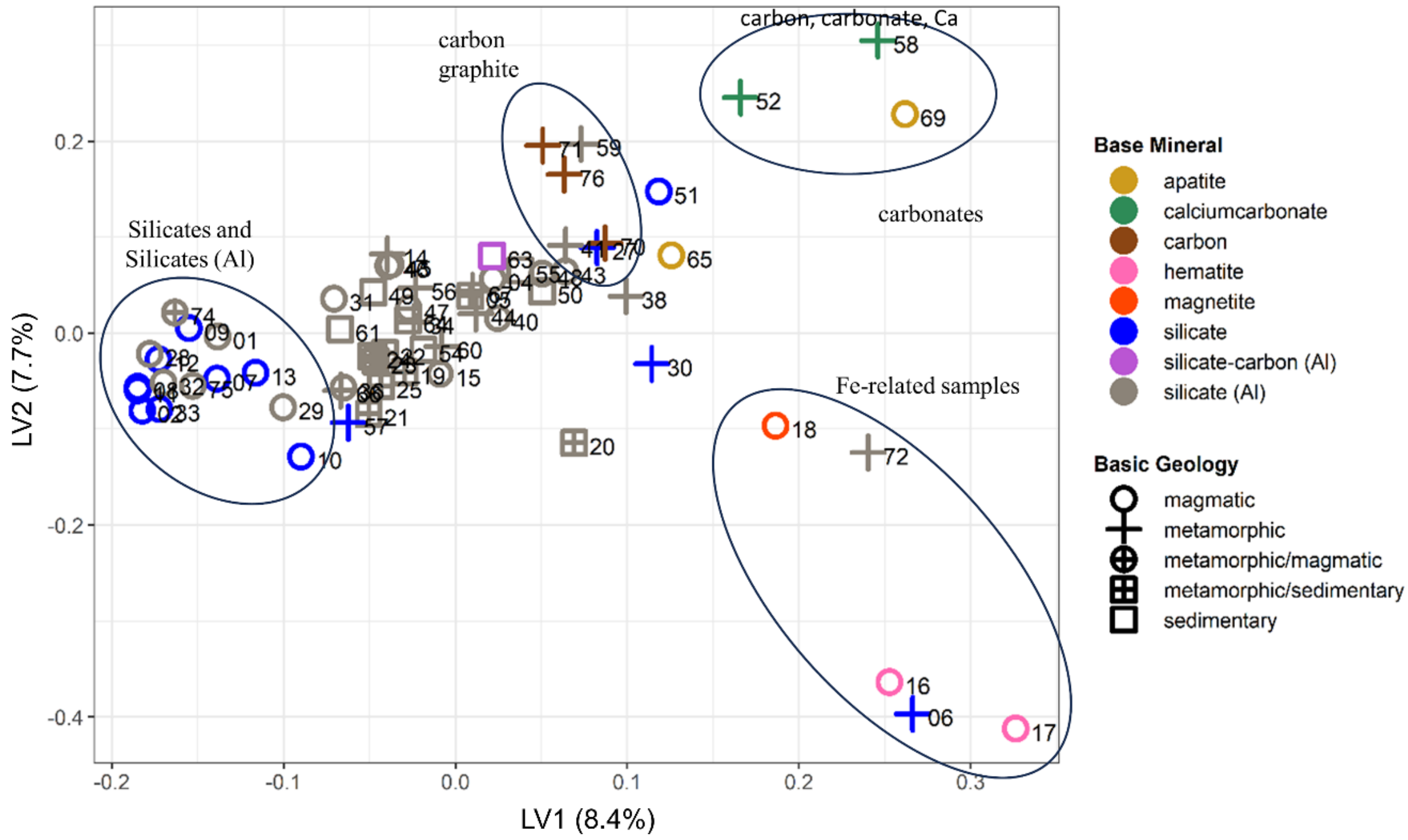

Hematites (16-17) and halleflint (06) contain high amounts of Fe, as do magnetite (18) and Ericssonit (72). In the carbon group, graphite (71,76) and eclogite (59) cluster. Eclogite is related to blueschist and may contain graphite/carbon hence the grouping. Silicates with/without Al could not be fully discriminated by the HPLS model.

**Fig. S18** Scores plot of LV1 and LV2 for the VIP-refined HPLS model of all five spectral data sets with clusters marked by mineral information:

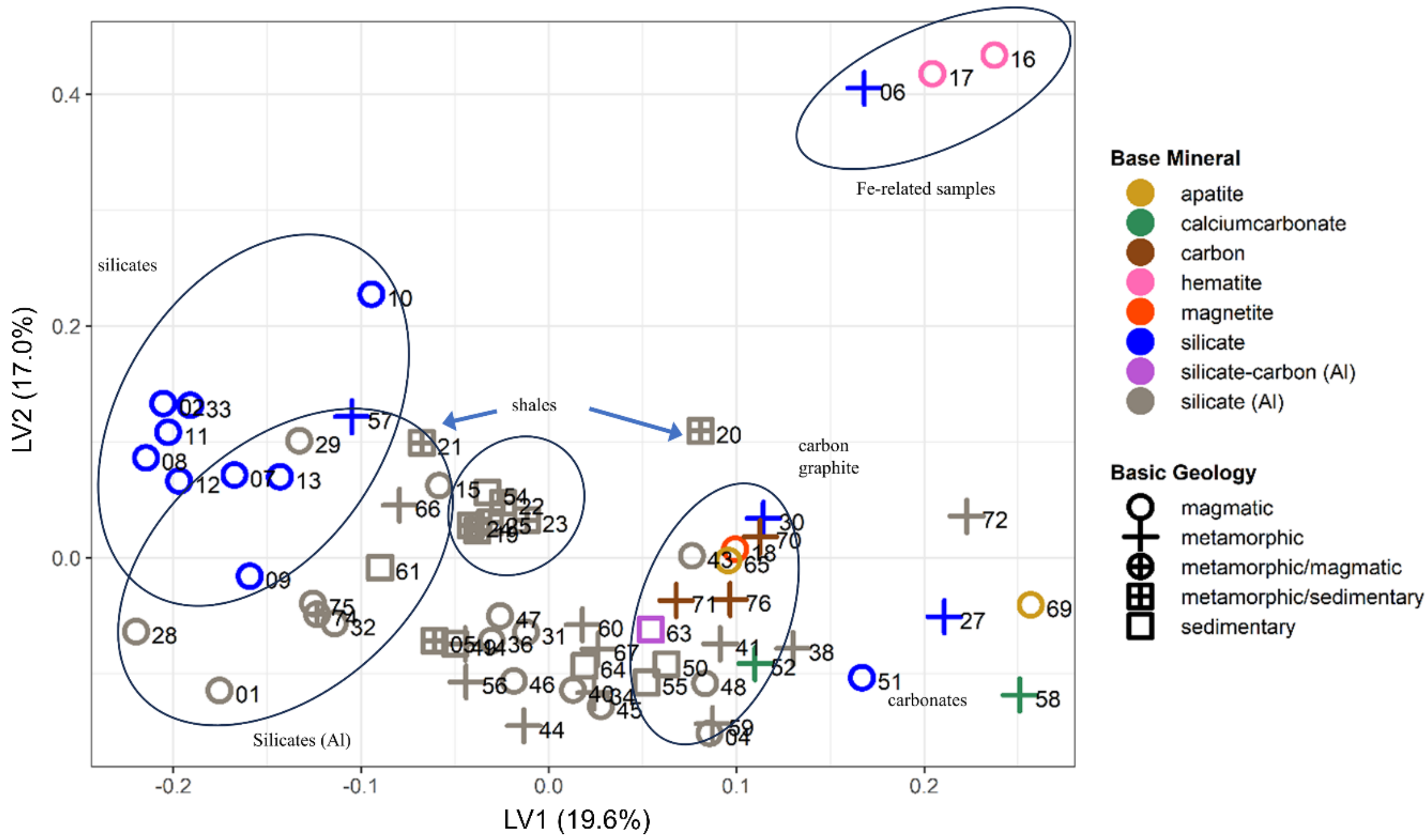

Slightly different clustering in the scores plot of the new HPLS model generated after reducing the dimensions of the original HPLS using VIP. In the VIP-HPLS model, the silicates with aluminium were discriminated from the silicates without aluminium.

# TABLES

**Table S1.** General information of the samples:

| Sample ID | Sample name | Location | Site | Province | Country |
|---|---|---|---|---|---|
| **01** | pegmatite | Ålöterna Ö.ED | Ålöterna | Småland | Sweden |
| **02** | rose quartz | Godegård | Lidbacken | Östergötland | Sweden |
| **03** | Smålandic granite | Västervik | Västervik | Kalmar | Sweden |
| **04** | diabase | Ekön, Gryt | Ekön | Östergötland | Sweden |
| **05** | shale | Persbergshyttan | Persbergshyttan | Värmland | Sweden |
| **06** | halleflint | Långban | Långban | Värmland | Sweden |
| **07** | rose quartz | - | ND | ND | ND |
| **08** | milky quartz | - | ND | ND | ND |
| **09** | smoky quartz | Björnsmåla | Björnsmåla | Östergötland | Sweden |
| **10** | smoky quartz | Okänd lokalitet | ND | ND | ND |
| **11** | rose quartz | SGU | SGU | ND | ND |
| **12** | rock crystal | Falerum | Falerum | Östergötland | Sweden |
| **13** | blue quartz | - | ND | ND | ND |
| **14** | gneiss | - | ND | ND | ND |
| **15** | porphyry | Långban, Värmland | Långban | Värmland | Sweden |
| **16** | hematite | Långban, Värmland | Långban | Värmland | Sweden |
| **17** | hematite | Okänd proviniens | ND | ND | ND |
| **18** | magnetite | Nordmaling | Nordmaling | Västerbotten | Sweden |
| **19** | shale | Åseleälven | Åseleälven | Ångermanland | Sweden |
| **20** | shale | Ormsjö, Dorotea | Ormsjö | Västerbotten | Sweden |
| **21** | shale | Vilhelmina | Vilhelmina | Västerbotten | Sweden |
| **22** | shale | Risbäck Dorotea | Risbäck | Västerbotten | Sweden |
| **23** | shale | - | ND | ND | ND |
| **24** | shale | Drömgruvan, Kolmården | Drömgruvan | Östergötland | Sweden |
| **25** | shale | Långban, Värmland | Långban | Värmland | Sweden |
| **26** | biotite | Hannäs bergrum, Sprängmassan | Hannäs bergrum | Östergötland | Sweden |
| **27** | actinolite | Falerum, Småland, Okänt ursprung | Falerum | Småland | Sweden |
| **28** | albite | Långban, Värmland | Långban | Värmland | Sweden |
| **29** | blue quartz | Lilla Älgsjöbrottet, Kolmården | Lilla Älgsjöbrottet | Östergötland | Sweden |
| **30** | talc | Hulta V.Ed | Hulta | Småland | Sweden |
| **31** | albite | Lidbacken, Godegård, Östergötland | Lidbacken | Östergötland | Sweden |
| **32** | orthoclase | Drömgruvan, Östergötland | Drömgruvan | Östergötland | Sweden |
| **33** | quartz | Ransäterhöjden, Värmland | Ransäterhöjden | Värmland | Sweden |

| 34 | muscovite | Lidbacken, Godegård, Östergötland | Lidbacken | Östergötland | Sweden |
|---|---|---|---|---|---|
| 35 | Amethyst | Hålsjöberg, Värmland | Hålsjöberg | Värmland | Sweden |
| 36 | orthoclase | Hyulta, Småland | Hulta | Småland | Sweden |
| 37 | kyanite quartzite | SGU | SGU | ND | ND |
| 38 | biotite | Lavergruvan, Älvsbyn s:n, Norrbotten | Lavergruvan | Norrbotten | Sweden |
| 39 | biotite | Långban, Värmland | Långban | Värmland | Sweden |
| 40 | greenstone | B Gård Os-Fältet Norway | Gård-Os | Hordaland | Norway |
| 41 | soapstone | Island | Island | ND | Island |
| 42 | greenstone | - | ND | ND | ND |
| 43 | pumice stone | Mora, W-län | Mora | Dalarna | Sweden |
| 44 | mica slate | Nordingrå | Nordingrå | Ångermanland | Sweden |
| 45 | diabase | Island | Island | ND | Island |
| 46 | rapakivi granite | Nordingå | Nordingå | Ångermanland | Sweden |
| 47 | uff | - | ND | ND | ND |
| 48 | gabbro | Grythyttan | Grythyttan | Västmanland | Sweden |
| 49 | sandstone | Kittelfjäll, V-bottens län | Kittelfjäll | Västerbotten | Sweden |
| 50 | greywacke | Champagne, Frankrike | Champagne | Champagne | France |
| 51 | olivine | - | ND | ND | ND |
| 52 | chalk | Nordingrå | Nordingrå | Ångermanland | Sweden |
| 53 | phyllite | SE om Hjellar Os Norway | Hjellar-Os | Hordaland | Norway |
| 54 | sandstone | Garpenberg | Garpenberg | Dalarna | Sweden |
| 55 | greenschist | Garpenberg, Norra | Garpenberg | Dalarna | Sweden |
| 56 | sericite quartzite | - | ND | ND | ND |
| 57 | quartzite | - | ND | ND | ND |
| 58 | marble | Vallintjåkko, Tärna, Västerbotten | Vallintjåkko | Västerbotten | Sweden |
| 59 | eclogite | Bredtorp, Tryserum | Tryserum | Småland | Sweden |
| 60 | quartzite | Degerhamn, Öland | Degerhamn | Öland | Sweden |
| 61 | sandstone | Boda kalkbrott, Dalarna | Boda kalkbrott | Dalarna | Sweden |
| 62 | Alum Shale | Zinkgruvan | Zinkgruvan | ND | Sweden |
| 63 | Alum Shale | Hendry Kiirunavaara | Kiirunavaara | Norrbotten | Sweden |
| 64 | sandstone | Bäckfall, Ödesäng, Tryserum | Tryserum | Småland | Sweden |
| 65 | apatite-hematite | Åläng, Skrickerum | Åläng | Östergötland | Sweden |
| 66 | halleflint | - | ND | ND | ND |
| 67 | amphibolite | Sovjetunionen | ND | ND | Russia |

| **68** | anthracite | C Sätragruvan, Östergötland | Sätragruvan | Östergötland | Sweden |
|---|---|---|---|---|---|
| **69** | apatite | - | ND | ND | ND |
| **70** | graphite | Hässelkulla västra, Örebro | Hässelkulla västra | Närke | Sweden |
| **71** | anthracite | Åsen i Skåne | Åsen | Skåne | Sweden |
| **72** | ericssonite | Utö, Stockholms Län | Utö | Södermanland | Sweden |
| **73** | fossilised wood | - | ND | ND | ND |
| **74** | blue tourmaline | Skrammelfallsgruvan, Halvarsbenning, Norberg, U-län | Skrammelfallsgruvan | Västmanland | Sweden |
| **75** | microcline | | ND | ND | ND |
| **76** | graphite | | ND | ND | ND |

**Table S2.** Geological and chemical information of the samples:

| Sample ID | General Name | Mineral Information | Basic Geology | Base Mineral | Chemistry |
|---|---|---|---|---|---|
| 01 | pegmatite | silicate | magmatic | silicate (Al) | $SiO_2$ |
| 02 | quartz | quartz | magmatic | silicate | Ti-Mg-$SiO_2$ |
| 03 | granite | feldspar, quartz-mica, hornblende, pyroxene | magmatic | silicate (Al) | |
| 04 | diabase | Plagioclase, pyroxene | magmatic | silicate (Al) | $SiO_2$ |
| 05 | shale | quartz and feldspar | metamorphic /sedimentary | silicate (Al) | |
| 06 | halleflint | quartz and feldspar | metamorphic | silicate | |
| 07 | quartz | quartz | magmatic | silicate | Ti-Mg-$SiO_2$ |
| 08 | quartz | quartz | magmatic | silicate | $SiO_2$ |
| 09 | quartz | quartz | magmatic | silicate | $SiO_2$ |
| 10 | quartz | quartz | magmatic | silicate | $SiO_2$ |
| 11 | quartz | quartz | magmatic | silicate | Ti-Mg-$SiO_2$ |
| 12 | quartz | quartz | magmatic | silicate | $SiO_2$ |
| 13 | quartz | quartz | magmatic | silicate | $SiO_2$ |
| 14 | gneiss | quartz and feldspar mica | metamorphic | silicate (Al) | |
| 15 | porphyry | quartz and feldspar hornblende | magmatic | silicate (Al) | $SiO_2$ |
| 16 | hematite | hematite | magmatic | hematite | $\alpha$-$Fe_2O_3$ |
| 17 | hematite | hematite | magmatic | hematite | $\alpha$-$Fe_2O_3$ |
| 18 | magnetite | magnetite | magmatic | magnetite | $Fe_3O_4$ |
| 19 | shale | quartz and feldspar | metamorphic /sedimentary | silicate (Al) | |
| 20 | shale | quartz and feldspar | metamorphic /sedimentary | silicate (Al) | |
| 21 | shale | quartz and feldspar | metamorphic /sedimentary | silicate (Al) | |
| 22 | shale | quartz and feldspar | metamorphic /sedimentary | silicate (Al) | |
| 23 | shale | quartz and feldspar | metamorphic /sedimentary | silicate (Al) | |
| 24 | shale | quartz and feldspar | metamorphic /sedimentary | silicate (Al) | |
| 25 | shale | quartz and feldspar | metamorphic /sedimentary | silicate (Al) | |
| 26 | biotite | mica | metamorphic | silicate (Al) | $K(Mg, Fe)_3AlSi_3O_{10}(F, OH)_2$ |
| 27 | aktinolite | inosilicate | metamorphic | silicate | $Ca_2(Mg, Fe_2^{+})_5Si_8O_{22}(OH)_2$ |
| 28 | albite | feldspar, plagioclase, tectosilicate | magmatic | silicate (Al) | $NaAlSi_3O_8$ |
| 29 | quartz | quartz | magmatic | silicate (Al) | $SiO_2$ |
| 30 | talc | talc | metamorphic | silicate | $Mg_3Si_4O_{10}(OH)_2$ |
| 31 | albite | feldspar, plagioclase, tectosilicate | magmatic | silicate (Al) | $NaAlSi_3O_8$ |
| 32 | orthoclase | potassium feldspar | magmatic | silicate (Al) | $K\text{-}Al\text{-}SiO_2$ |
| 33 | quartz | quartz | magmatic | silicate | $SiO_2$ |
| 34 | muscovite | mica | metamorphic | silicate (Al) | $KAl_2(AlSi_3O_{10})(OH)_2$ |
| 35 | quartz | quartz | magmatic | silicate (Al) | $SiO_2$ |
| 36 | orthoclase | potassium feldspar | magmatic | silicate (Al) | $K\text{-}Al\text{-}SiO_2$ |
| 37 | quartzite | quartz | metamorphic | silicate (Al) | $Al\text{-}SiO_2$ |
| 38 | biotite | mica | metamorphic | silicate (Al) | $K(Mg, Fe)_3AlSi_3O_{10}(F, OH)_2$ |
| 39 | biotite | mica | metamorphic | silicate (Al) | $K(Mg, Fe)_3AlSi_3O_{10}(F, OH)_2$ |
| 40 | greenstone | chlorite, epidote, albite, pyroxene, olivine och biotite | magmatic | silicate (Al) | |
| 41 | soapstone | magnesium silicate/chlorite | metamorphic | silicate (Al) | $Mg\ SiO_2,(Mg,Fe)_3(Si,Al)_4O_{10}(OH)_2\cdot(Mg,Fe)_3(OH)$ |
| 42 | greenstone | chlorite, epidote, albite, pyroxene, olivine och biotite | magmatic | silicate (Al) | |
| 43 | pumice | felsite, rhyolite or basalt | magmatic | silicate (Al) | |
| 44 | slate | feldspar, quartz and mica | metamorphic | silicate (Al) | |
| 45 | diabase | plagioclase, pyroxene | magmatic | silicate (Al) | $SiO_2$ |
| 46 | granite | feldspar, quartz-mica, hornblende and pyroxene | magmatic | silicate (Al) | |
| 47 | Tuff | high-silica rhyolitic ash to low-silica basaltic ash | magmatic | silicate (Al) | |

| 48 | gabbro | pyroxene, plagioclase, and minor amounts of amphibole and olivine. | magmatic | silicate (Al) | |
|---|---|---|---|---|---|
| 49 | sandstone | quartz, feldspar | sedimentary | silicate (Al) | $SiO_2$ |
| 50 | greywacke | quartz, feldspar | sedimentary | silicate (Al) | |
| 51 | olivine | magnesium-iron silicate | magmatic | silicate | $(Fe,Mg)_2SiO_4$ |
| 52 | chalk | calcium carbonate | metamorphic | calciumcarbonate | $CaCO_3$ |
| 53 | phyllite | quartz, sericite and chlorite | metamorphic | silicate (Al) | $SiO_2$ |
| 54 | sandstone | quartz, feldspar | sedimentary | silicate (Al) | $SiO_2$ |
| 55 | shale | quartz, feldspar | sedimentary | silicate (Al) | |
| 56 | quartzite | quartz, sericite | metamorphic | silicate (Al) | |
| 57 | quartzite | quartz | metamorphic | silicate | $SiO_2$ |
| 58 | marble | calcium carbonate | metamorphic | calciumcarbonate | $CaCO_3$ |
| 59 | eclogite | pyroxene, hornblende and garnet | metamorphic | silicate (Al) | |
| 60 | quartzite | quartz | metamorphic | silicate (Al) | $SiO_2$ |
| 61 | sandstone | quartz, feldspar | sedimentary | silicate (Al) | $SiO_2$ |
| 62 | shale | mica | sedimentary | silicate-carbon (Al) | $SiO_2$-C-$CO_3$ |
| 63 | shale | mica | sedimentary | silicate-carbon (Al) | $SiO_2$-C-$CO_3$ |
| 64 | sandstone | quartz, feldspar | sedimentary | silicate (Al) | $SiO_2$ |
| 65 | apatite-hematite | apatite | magmatic | apatite | $Ca_{10}(PO_4)_6(OH)_2$, $Ca_{10}(PO_4)_6F$ - Fe |
| 66 | halleflint | quartz and feldspar | metamorphic | silicate (Al) | |
| 67 | amphibolite | hornblende plagioclase (feldspar) Inosilicate | metamorphic | silicate (Al) | |
| 68 | anthracite | pre grafit | metamorphic | carbon | C |
| 69 | apatite | apatite | magmatic | apatite | $Ca_{10}(PO_4)_6(OH)_2$, $Ca_{10}(PO4)_6F$ |
| 70 | graphite | graphite | metamorphic | carbon | C |
| 71 | anthracite | bituminous coal to graphite | metamorphic | carbon | C |
| 72 | ericssonite | sorosilicates - talc | metamorphic | silicate (Al) | $BaMn_{22}+Fe_3+(Si_2O_7)O(OH)$ |
| 73 | fossilised wood | quartz - carbon? | metamorphic | silicate (Al) | C-$SiO_2$ |
| 74 | tourmaline | aluminium-borosilicates | metamorphic/magmatic | silicate (Al) | |
| 75 | microcline | feldspar | magmatic | silicate (Al) | $KAlSi_3O_8$ |
| 76 | graphite | graphite | metamorphic | carbon | C |

**Table S3.** Total and per component explained variation of the ASD-NIR PLS model:

| ASD-NIR PLS model | | | | | |
|---|---|---|---|---|---|
| **LV** | **R2X** | **R2Xcum** | **R2Y** | **R2Ycum** | **RMSE** |
| 1 | 60.7 | 60.7 | 8.4 | 8.4 | 8.2 |
| 2 | 22.2 | 82.9 | 6.3 | 14.7 | 8.1 |
| 3 | 6.3 | 89.2 | 6.8 | 21.4 | 8.0 |
| 4 | 3.7 | 92.8 | 4.5 | 25.9 | 8.2 |
| 5 | 2.5 | 95.3 | 5.6 | 31.5 | 7.8 |
| 6 | 2.0 | 97.3 | 3.7 | 35.2 | 7.6 |
| 7 | 0.5 | 97.9 | 4.3 | 39.5 | 7.5 |
| 8 | 0.3 | 98.1 | 6.5 | 46.0 | 7.5 |
| 9 | 0.3 | 98.4 | 4.3 | 50.2 | 7.4 |
| 10 | 0.8 | 99.1 | 0.9 | 51.2 | 7.4 |
| 11 | 0.2 | **99.4** | 1.4 | **52.5** | 7.4 |

**Table S4.** Total and per component explained variation of the Micro-NIR PLS model:

| Micro-NIR PLS model | | | | | |
|---|---|---|---|---|---|
| **LV** | **R2X** | **R2Xcum** | **R2Y** | **R2Ycum** | **RMSE** |
| 1 | 76.1 | 76.1 | 6.7 | 6.7 | 8.2 |
| 2 | 9.9 | 86.0 | 5.7 | 12.4 | 8.1 |
| 3 | 6.9 | 92.9 | 2.0 | 14.4 | 8.1 |
| 4 | 5.1 | 98.0 | 1.1 | 15.5 | 8.2 |
| 5 | 1.1 | 99.1 | 1.7 | 17.2 | 8.2 |
| 6 | 0.4 | 99.5 | 2.7 | 19.9 | 8.3 |
| 7 | 0.2 | **99.7** | 3.8 | **23.7** | 8.3 |

**Table S5.** Total and per component explained variation of the Bruker-Raman PLS model:

| Bruker-Raman PLS model | | | | | |
|---|---|---|---|---|---|
| **LV** | **R2X** | **R2Xcum** | **R2Y** | **R2Ycum** | **RMSE** |
| 1 | 18.9 | 18.9 | 14.6 | 14.6 | 8.1 |
| 2 | 14.3 | 33.2 | 10.2 | 24.8 | 7.9 |
| 3 | 12.4 | 45.5 | 6.8 | 31.6 | 7.6 |
| 4 | 8.1 | 53.6 | 7.1 | 38.7 | 7.6 |
| 5 | 7.9 | 61.5 | 3.4 | 42.1 | 7.7 |
| 6 | 5.6 | 67.1 | 4.1 | 46.2 | 7.6 |
| 7 | 3.7 | 70.8 | 4.6 | 50.8 | 7.6 |
| 8 | 4.5 | 75.4 | 3.8 | 54.6 | 7.3 |
| 9 | 2.3 | 77.6 | 4.0 | 58.6 | 7.4 |
| 10 | 2.3 | 80.0 | 3.0 | 61.6 | 7.5 |
| 11 | 2.0 | 81.9 | 2.8 | 64.4 | 7.6 |
| 12 | 1.5 | 83.4 | 3.7 | 68.1 | 7.7 |
| 13 | 1.0 | 84.4 | 4.4 | 72.5 | 7.8 |
| 14 | 1.5 | 85.9 | 2.8 | 75.3 | 7.8 |
| 15 | 1.9 | **87.8** | 2.0 | **77.3** | 7.7 |

**Table S6.** Total and per component explained variation of the i-Raman PLS model:

| i-Raman PLS model | | | | | |
|---|---|---|---|---|---|
| **LV** | **R2X** | **R2Xcum** | **R2Y** | **R2Ycum** | **RMSE** |
| 1 | 50.9 | 50.9 | 13.2 | 13.2 | 8.1 |
| 2 | 25.7 | 76.6 | 4.5 | 17.7 | 8.1 |
| 3 | 8.6 | 85.2 | 7.4 | 25.1 | 8.0 |
| 4 | 6.0 | 91.2 | 2.9 | 28.1 | 8.2 |
| 5 | 4.7 | 95.9 | 3.7 | 31.7 | 8.4 |
| 6 | 0.9 | 96.8 | 7.8 | 39.6 | 8.1 |
| 7 | 0.6 | **97.4** | 3.6 | **43.1** | 8.3 |

**Table S7.** Total and per component explained variation of the HPLS model including all five spectral data sets (ASD-NIR, Micro-NIR, Bruker-Raman, i-Raman, XRF):

| HPLS model of NIR, FT-Raman and XRF data | | | | | |
|---|---|---|---|---|---|
| **LV** | **R2X** | **R2Xcum** | **R2Y** | **R2Ycum** | **RMSE** |
| 1 | 8.4 | 8.4 | 35.4 | 35.4 | 317.1 |
| 2 | 7.7 | 16.1 | 15.1 | 50.5 | 293.9 |
| 3 | 6.7 | 22.8 | 7.7 | 58.2 | 285.8 |
| 4 | 6.1 | 29.0 | 7.3 | 65.5 | 269.0 |
| 5 | 5.1 | 34.1 | 3.6 | 69.1 | 270.4 |
| 6 | 4.2 | 38.3 | 5.1 | 74.1 | 269.8 |
| 7 | 4.5 | 42.8 | 2.1 | 76.2 | 270.4 |
| 8 | 4.0 | 46.7 | 2.9 | 79.1 | 272.3 |
| 9 | 5.9 | 52.7 | 1.2 | 80.3 | 267.3 |
| 10 | 4.6 | 57.3 | 1.5 | 81.8 | 264.2 |
| 11 | 4.2 | 61.5 | 1.2 | 82.9 | 257.6 |
| 12 | 4.5 | 66.1 | 0.7 | 83.6 | 255.1 |
| 13 | 3.3 | 69.3 | 1.2 | 84.8 | 253.1 |
| 14 | 3.8 | **73.1** | 0.6 | **85.4** | 252.6 |

**Table S8.** Total and per component explained variation of the HPLS model dimensionally reduced using VIP method including all five spectral data sets (ASD-NIR, Micro-NIR, Bruker-Raman, i-Raman, XRF):

| VIP-HPLS model of NIR, FT-Raman and XRF data | | | | | |
|---|---|---|---|---|---|
| **LV** | **R2X** | **R2Xcum** | **R2Y** | **R2Ycum** | **RMSE** |
| 1 | 19.58 | 19.58 | 41.37 | 41.37 | 240.9 |
| 2 | 17.01 | 36.59 | 9.6 | 50.97 | 226.5 |
| 3 | 10.51 | 47.1 | 5.55 | 56.52 | 220.9 |
| 4 | 9.66 | 56.76 | 4.26 | 60.78 | 212.6 |
| 5 | 8.64 | 65.4 | 1.9 | 62.68 | 218.4 |
| 6 | 9.55 | 74.95 | 1.32 | 64 | 211.4 |
| 7 | 4.39 | 79.35 | 1.83 | 65.83 | 213.5 |
| 8 | 7.19 | 86.53 | 0.46 | 66.28 | 213.2 |
| 9 | 5.25 | 91.78 | 0.7 | 66.99 | 209.9 |
| 10 | 3.16 | 94.94 | 0.65 | 67.63 | 210.6 |
| 11 | 1.5 | 96.44 | 0.37 | 68 | 216.4 |
| 12 | 1.94 | 98.38 | 0.35 | 68.35 | 214.1 |
| 13 | 1.25 | 99.63 | 0.23 | 68.58 | 212.6 |
| 14 | 0.37 | **100** | 0.16 | **68.74** | 214.8 |